\documentclass[a4paper,11pt]{article}
\pdfoutput=1 

\usepackage{jcappub} 
\usepackage{lineno}

\usepackage[normalem]{ulem}
\usepackage[utf8]{inputenc}
\usepackage{enumerate}
\usepackage{amsmath, amssymb, amsthm, graphicx, epsfig, fancyhdr,epsfig, slashed}
\usepackage[mathscr]{euscript}
\let\oldcdot\cdot
\usepackage{breqn}
\let\cdot\oldcdot
\usepackage{epsfig}  
\usepackage{graphicx}   
\usepackage{slashed}       
\usepackage{tikz}
\usepackage{subcaption} 
\usepackage{url}
\usepackage{color,soul} 
\usepackage{xcolor}
\usepackage{multirow} 
\usepackage{comment}
\usepackage{mathtools}
\usepackage{bm}
\usepackage{hyperref}
\usepackage[all]{hypcap}
\hypersetup{  
    colorlinks=true,
    linkcolor=blue,
    filecolor=red,      
    urlcolor=cyan,
    citecolor=blue,
    }

\arxivnumber{} 

\newcommand{\be}{\begin{eqnarray}}
\newcommand{\ee}{\end{eqnarray}}
\newcommand{\deltat}{\overline{T}_{21}}
\newcommand{\ts}{T_s}
\newcommand{\tk}{T_k}
\newcommand{\tgamma}{T_{\gamma}}

\newcommand{\xalpha}{x_{\alpha}}
\newcommand{\sfrd}{\dot{\rho}_{\star}}
\newcommand{\nb}{n_{\!_{B}}}

\begin{document}
\title{\sf Revisiting primordial magnetic fields through 21-cm physics: Bounds and forecasts}

\author[a]{Arko Bhaumik,}
\author[a]{Debarun Paul,}
\author[a,b]{Supratik Pal}

\affiliation[a]{\,Physics and Applied Mathematics Unit, Indian Statistical Institute, 203 B.T. Road,\\ Kolkata 700108, India}
\affiliation[b]{\,Technology Innovation Hub on Data Science, Big Data Analytics and Data Curation, Indian Statistical Institute, 203 B.T. Road, Kolkata 700108, India}

\emailAdd{arkobhaumik12@gmail.com}
\emailAdd{debarun31paul@gmail.com}
\emailAdd{supratik@isical.ac.in}

\abstract{Primordial magnetic fields (PMFs) may significantly influence 21-cm physics via two mechanisms: (i) magnetic heating of the intergalactic medium (IGM) through ambipolar diffusion (AD) and decaying magnetohydrodynamic turbulence (DT), (ii) impact on the star formation rate density (SFRD) through small-scale enhancement of the matter power spectrum. In this analysis, we integrate both of these effects within a unified analytical framework and use it to determine upper bounds on the parameter space of a nearly scale-invariant non-helical PMF in the light of the global 21-cm signal observed by EDGES. Our findings reveal that the joint consideration of both effects furnishes constraints of the order $B_0\lesssim\mathcal{O}(10^{-2})$ nG on the present-day magnetic field strength, which are considerably tighter compared to earlier analyses. We subsequently explore the prospects of detecting such a magnetized 21-cm power spectrum at the upcoming SKA-Low mission. For the relevant parameters of the PMF ($B_0$ and $n_{\!_{B}}$) and the excess radio background ($\xi$), SNR estimation and Fisher forecast analysis indicate that it may be possible to constrain these three parameters with relative $1\sigma$ uncertainties $\lesssim10\%$ and an associated SNR $\gtrsim10$ at SKA-Low. This also leads to possible correlations among these three parameters, thus revealing intriguing trends of interplay among the various physical processes involved.
}

\maketitle

\section{Introduction}
\label{sec:intro}

The presence of very weak magnetic fields on cosmic scales has been hinted at for decades by several independent observations. In particular, it is currently known that magnetic fields typically of strength $10^{-15}\:\textrm{G}<B_0<10^{-9}\:\textrm{G}$ at the present day may permeate the voids of the intergalactic medium (IGM) across length scales of $\mathcal{O}(1)\:\rm{Mpc}$. The $\mathcal{O}(\rm{nG})$ upper bound is furnished by Faraday rotation measures of extragalactic sources \cite{faraday_pmf_bound_1,faraday_pmf_bound_2,faraday_pmf_bound_3,faraday_pmf_bound_4} and cosmic microwave background (CMB) constraints \cite{cmb_pmf_bound_1,cmb_pmf_bound_2,cmb_pmf_bound_3}, while the lower bound results from the null detection of secondary (GeV) $\gamma$-ray cascades triggered in the IGM by primary (TeV) $\gamma$-emissions from distant blazars \cite{blazar_bound_1,blazar_bound_2,blazar_bound_3}. The inadequacy of bottom-up astrophysical dynamo mechanisms \cite{dynamo_rev_1,dynamo_rev_2,Rev_Beck_2019,igmf_astro} when it comes to the explanation of such large-scale volume-filling magnetic fields indicates a possible primordial origin of these fields, which may be produced, for example, via a wide range of inflationary dynamics \cite{ratra,martin_yokoyama,sawtooth,bounce_fields,resolve_3,grav_coup_1,grav_coup_2,grav_coup_3,helical_1,helical_2,helical_3} or due to phase transitions in the early Universe \cite{pmf_phase_1,pmf_phase_2,pmf_phase_3,pmf_phase_4,pmf_phase_5,pmf_phase_6,pmf_phase_7,pmf_phase_8,pmf_phase_9,pmf_phase_10} (for brief reviews on both, see Refs.~\cite{magnetogen_rev_1,magnetogen_rev_2}). Beyond these two major windows, there may have existed several other exotic mechanisms in the early Universe that could possibly have led to the generation of cosmic magnetic fields, \textit{e.g.} based on cosmological density perturbations \cite{Ichiki:2006cd,Naoz:2013wla}, primordial black holes \cite{Safarzadeh:2017mdy,Hooper:2022cvr,Papanikolaou:2023nkx,Papanikolaou:2023cku}, nonlinear electrodynamics \cite{nld} and extra-dimensional physics \cite{extradim}, \textit{etc}. Owing to its stochastic nature, such a  primordial magnetic field (PMF) is typically characterized by its power spectrum, whose amplitude is governed by the magnetic energy density smoothed on a given scale at present time ($A_B$), and the scale-dependence is regulated by the magnetic spectral index ($n_{\!_{B}}$). Constraints on $A_B$ can be directly translated to bounds on the present magnetic field strength if its redshift-evolution is properly incorporated, whereas currently available datasets do not put precise constraints on $n_{\!_{B}}$, which is usually chosen to be a free parameter while modelling various cosmological effects of PMFs across the various cosmic epochs. 

Such ambient magnetic fields are naturally expected to leave some imprints on the redshifted neutral hydrogen (HI) 21-cm signal. Being a treasure trove of information as far as the physics of the dark ages and the cosmic dawn is concerned, the 21-cm observational window could be a potentially important tool in constraining the parameters of the PMF power spectrum as well. A stochastic PMF could be either helical or non-helical in nature. The latter being the minimal scenario from a primordial magnetogenetic point of view \cite{magnetogen_rev_1,magnetogen_rev_2}, we stick to the non-helical PMF only in this work. The key processes through which such a non-helical PMF could affect the 21-cm signal may be classified broadly under two categories, namely IGM heating effects \cite{heating_1,heating_2,heating_3,heating_4,heating_5} and possible impact on the formation of large scale structure (LSS) \cite{structure_1,structure_2,structure_3,structure_4}, both of which are well-studied from theoretical perspectives. The former results from the dissipation of magnetic energy through ambipolar diffusion (AD) and decaying magnetohydrodynamic turbulence (DT) which may heat up the kinetic sector of the gaseous IGM, while the latter can be traced to the small-scale enhancement of matter power induced by a stochastic PMF. The Experiment to Detect the Global EoR Signature (EDGES)~\cite{Bowman:2018yin} has reported an intense depth in the global 21-cm absorption signal $\deltat(z)$ which is in 3.8$\sigma$ tension with the prediction of the vanilla $\Lambda$CDM scenario~\cite{Barkana:2018lgd}. Following this detection, the impact of magnetic IGM heating (via AD \& DT) on the 21-cm dip was considered in Ref.~\cite{Minoda:2018gxj}, where an upper limit of $B_0\lesssim\mathcal{O}(0.1)$ nG was obtained based on the requirement that the 21-cm signal be in absorption (as observed by EDGES) rather than in emission in an otherwise $\Lambda$CDM background. Magnetic effects on the IGM temperature and resulting PMF constraints in the light of the EDGES detection have subsequently been investigated for various other non-minimal physical scenarios as well, \textit{e.g.} baryon-dark matter (DM) interactions \cite{Bhatt:2019lwt,Bera:2020jsg}, helical PMF spectra \cite{Natwariya:2020mhe}, and the presence of excess radio background \cite{Natwariya:2020ksr}. One may readily observe that each of these scenarios accommodates some specific mechanism via which the IGM can be sufficiently cooled so as to allow $\deltat(z)=-500^{+200}_{-500}$ mK in the redshift range $16\lesssim z \lesssim19$ as observed by EDGES.

However, another crucial feature which needs to be taken into account while constraining a PMF on the basis of the existing global 21-cm data is its role in the formation of structure, which is expected to play a significant role in the redshift evolution of $\deltat(z)$ by directly influencing the star formation history of our Universe. While such effects are well explored in the literature via both analytical and simulation-based approaches \cite{2019FrASS...6....7K,Sharda:2020aio,Koh:2021gjt,Prole:2022pvv,2023MNRAS.520...89W,Adi:2023qdf,2024arXiv240305672S}, a detailed investigation of their role in the context of the EDGES signal is lacking in the aforementioned studies, which have focused mostly on the magnetic heating effects in the IGM. The latter results primarily in a vertical shift of the amplitude of $\deltat(z)$, which has been the cornerstone of  EDGES-based analysis of PMFs so far. However, modification of the star formation rate density (SFRD) in presence of PMFs should have a significant impact on X-ray heating of the IGM, which is instrumental in determining the redshift of the 21-cm absorption minimum. More precisely, besides the aforementioned vertical change in its depth, one should also expect a lateral shift of the absorption trough towards higher (lower) redshifts if the SFRD is enhanced (suppressed) at earlier times in the presence of PMFs, which therefore provides an additional degree of freedom that may be used to constrain the PMF parameter space on the basis of existing data. 

In the present work, we first attempt to bridge this gap by integrating both magnetic heating effects (AD \& DT) and the impact of PMFs on the star formation history within a common analytical framework, and use it to derive PMF constraints based on the EDGES window. Because we focus on non-helical PMFs, we necessarily assume an excess radio background in order to achieve the required level of IGM cooling so as to be compatible with the EDGES bounds, as supported by the Absolute Radiometer for Cosmology, Astrophysics, and Diffuse Emission (ARCADE-2) \cite{fixsen2011arcade} and the Long Wavelength Array (LWA1) \cite{2012JAI.....150004T} experiments. If only magnetic heating effects are considered, the admissible upper bound on $B_0$ from EDGES for a nearly scale-invariant PMF can be significantly large and exceed $\mathcal{O}(1)$ nG \cite{Natwariya:2020ksr}, as magnetic heating can be readily compensated by the large parameter space of the excess radio background, all the while keeping the absorption trough comfortably within $16<z<19$. A magnetically modified SFRD model, however, changes the scenario by directly affecting the redshift at which the trough occurs, and as we find, allows deriving considerably tighter PMF constraints even in the presence of an excess radio background. Our analysis reveals that even for nearly scale-invariant non-helical PMFs, a stringent upper bound of $B_0\lesssim\mathcal{O}(10)$ pG may be achieved based primarily on this lateral shift, which appears to be significantly more sensitive to the PMF parameters than the dependence of the vertical depth of the signal on the PMF and the excess radio background combined. Furthermore, our analysis explicitly highlights the deviation from the simple $(1+z)^2$ redshift-scaling of the PMF amplitude owing to the dissipation of magnetic energy through AD and DT, which has also been pointed out earlier in \cite{Bera:2020jsg}. For the typical weak magnetic field strengths involved in this work, this relative deviation turns out to be significant and can lead to $\mathcal{O}(1)$ relative errors in the estimate of $B_0$. This has motivated us to model the scenario as a typical initial value problem where the magnetic field at recombination ($B_{\rm CMB}$) serves as the fundamental parameter, while the present day magnetic strength $B_0$ is a derived parameter which is subsequently obtained through a full numerical solution instead of being connected to $B_{\rm CMB}$ via an \textit{a priori} $B_{\rm CMB}=B_0(1+z)^{2}$ scaling. Thus, certain results in this work are presented in terms of $B_{\rm CMB}$ wherever deemed suitable.

In the second half of our work, we focus on the joint PMF and excess radio background parameter space that we find to be compatible with the global signal from EDGES, and explore the prospects of further constraints on the same based on 21-cm power spectrum data in the light of the upcoming Square Kilometre Array (SKA) \cite{Carilli:2004nx,2015aska.confE.174B,SKA:2018ckk,Weltman:2018zrl,2019arXiv191212699B}. In terms of HI 21-cm physics, the SKA is expected to be a transformative scientific facility of the ongoing century that would be able to shed light on hitherto unexplored epochs of our cosmic history, ranging from the dark ages at $30\lesssim z\lesssim100$ all the way up to the epoch of reionization (EoR) around $z\sim 4-12$. The scope of constraining cosmic magnetism on various scales with the SKA have been shown to be promising in numerous earlier works \cite{Gaensler:2006tj,Beck:2008qf,2016JApA...37...42R,structure_3,2020Galax...8...53H}. Thus, in addition to the presently available 21-cm window, it is imperative to assess the prospects of constraining the parameters of the scenario under consideration with future data from the SKA. For this purpose, we first calculate the signal-to-noise ratio (SNR) of the one-dimensional 21-cm power spectrum achievable by the SKA Low Frequency Array (SKA-Low) as a function of the two non-helical PMF parameters ($B_0$ and $n_{\!_{B}}$) and the ARCADE-2 modelling parameter ($\xi$) \cite{feng2018enhanced}. This is followed by a Fisher forecast analysis involving these three parameters, which allows us to estimate their projected $1\sigma$ errors at SKA-Low. For particular  fiducial values as allowed by the available parameter space from EDGES, our analysis reveals interesting trends for both the $1\sigma$ error forecasts and possible correlations from SKA-Low, which can be traced back to the interplay among the various physical effects induced by the PMF and the excess radio background.

The work is organised as follows. In Sec. \ref{sec:21cmsignal}, a brief outline of the theoretical framework for modelling the global $\deltat(z)$ signal in the presence of a non-helical PMF has been presented, by taking both IGM heating effects and the possible impact on star formation history into account. In Sec. \ref{sec:edgesbound}, we present our numerically obtained results for $\deltat(z)$ by including an excess radio background \textit{vis-\`{a}-vis} the observation by EDGES, thereby arriving at sets of upper bounds on $B_0$ and $\xi$ for two benchmark values of $n_{\!_{B}}=-2.75$ and $n_{\!_{B}}=-2.95$ (both representing nearly scale-invariant PMF power spectra). In Sec. \ref{sec:powspec}, we compute the 21-cm power spectra semi-analytically for suitable choices of the parameters that we find to be compatible with EDGES, and proceed to the subsequent estimation of the SNR and the Fisher forecast analysis in the light of SKA-Low in Sec. \ref{sec:noise}.  Finally, Sec. \ref{sec:conclusion} summarizes the key findings and presents a brief discussion on a few important future directions to explore.

Throughout this work, we assume $\mu_0=4\pi$ in Gaussian units for the magnetic permeability of vacuum, unless stated otherwise.

\section{The global 21-cm signal}
\label{sec:21cmsignal}

\subsection{21-cm signal in absence of magnetic field}
\label{subsec:vanilla21cmsignal}
The principal physical observable under investigation in any 21-cm experiment is the differential brightness temperature, denoted by $T_{21}$. The sky-averaged (global) differential brightness temperature can be expressed as ~\cite{FURLANETTO2006181,2012RPPh...75h6901P}
\be\label{eq:dTb}
\deltat (z) \simeq 27 x_{\rm HI}\left(\frac{\Omega_{b,0}h^2}{0.023}\right)\sqrt{\frac{0.15}{\Omega_{m,0} h^2}\frac{1+z}{10}}\left(1-\frac{\tgamma}{\ts}\right)\,\,\rm{mK}.
\ee
where $x_{\rm HI}$ is the neutral hydrogen fraction, and $\Omega_{m,0}$ ($\Omega_{b,0}$) represents the total (baryonic) matter abundance in the Universe at the present epoch. The spin temperature $\ts$ can be expressed as a weighted average of the background radiation temperature $T_\gamma$, the gas kinetic temperature $\tk$, and the Lyman-$\alpha$ temperature $T_\alpha$ as~\cite{1958PIRE...46..240F}
\be\label{eq:Ts}
\ts^{-1} = \frac{\tgamma^{-1} + x_k \tk^{-1} + \xalpha T_{\alpha}^{-1}}{1 + x_k + \xalpha}\:,
\ee
with $x_k$ and $\xalpha$ representing the collisional coupling coefficient and the Lyman-$\alpha$ coupling coefficient respectively. In our analysis, we consider $T_{\alpha} \approx \tk$, since frequent scattering among the Ly-$\alpha$ photons thermally equilibrate $T_\alpha$ and $\tk$~\cite{2012RPPh...75h6901P,1959ApJ...129..551F}. The coefficient $x_k$, incorporating three interaction channels (H-H, H-e, and H-p collisions), depends strongly on $\tk$ ~\cite{Kuhlen_2006,PhysRevD.74.103502}. On the other hand, $x_\alpha$ encapsulates the Lyman-$\alpha$ flux ($J_{\alpha}$)~\cite{PhysRevD.74.103502,Barkana_2005} and depends on the SFRD, which can be expressed as~\cite{Chen:2003gc,Pritchard_2006,Barkana:2004vb}
\be \label{eq:sfrd}
\sfrd(z)=-f_{\star}\Omega_{b,0}\rho_{c,0}(1+z)H(z)\dfrac{d}{dz}f_{\rm coll}(z)\:,
\ee
where $\rho_{c,0}\equiv\frac{3H_0^2}{8\pi G}$ is the critical density of the Universe at the present epoch and $f_{\star}$ represents the stellar-to-baryon ratio. Throughout this work, we have used $f_{\star}=0.01$ which aligns well with findings from radiation-hydrodynamic simulations of high-redshift galaxies~\cite{Schneider:2018xba,Wise_2014}. In the Press-Schechter (PS) formalism of spherical halo collapse~\cite{1974ApJ...187..425P}, the collapse function can be expressed as $f_{\rm coll}(z)=\textrm{erfc}\left(\frac{1.686(1+z)}{\sqrt{2}\sigma(R(z))}\right)$, where $\sigma^2$ is the mean-squared mass variance smoothed over a window function $W(k,R)$ of characteristic length scale $R(z)$ as
\be \label{eq:sigma}
\sigma^2(R(z))=\dfrac{1}{2\pi^2}\int\limits_0^\infty k^2P_m^{(\rm tot)}(k)|W(k,R(z))|^2\:.
\ee
Here, $P_m^{(\rm tot)}$ denotes the total matter power spectrum of the Universe at the present epoch. For the purpose of this study, we have chosen a standard top-hat window function of the form $W(k,R)=3\left[\sin(kR)-kR\cos(kR)\right]/(kR)^3$. Such a choice is reasonable as long as the matter power spectrum is sufficiently smooth and does not exhibit quick variations \cite{Schneider:2013ria}, which applies to our present scenario (\textit{viz.} Fig \ref{fig:mPk_and_SFRD}).

In order to model the global 21-cm signal, we additionally need the ionization histories of HI and HeI, whose ionizations fractions evolve as~\footnote{In this study, we have neglected HeII recombination as it does not significantly affect the CMB power spectra ~\cite{Seager:1999bc}.}~\cite{Seager:1999bc}

\begin{eqnarray}
\label{eq:xp_evol}
    {dx_{\rm p}\over dz} &= \left(x_{\rm e}x_{\rm p} n_{\rm \!_{H}} 
    \alpha_{\rm \!_{H}}
    - \beta_{\rm \!_{H}} (1-x_{\rm p})
    {\rm e}^{-h_{\!_{P}} \nu_{\!_{H2s}}/k_{\!_{B}} \tk}\right) \\
    &\times\quad{\left(1 + K_{\rm \!_{H}} \Lambda_{\rm \!_{H}} n_{\rm \!_{H}}(1-x_{\rm p})\right)
    \over H(z)(1+z)\left(1+K_{\rm \!_{H}} (\Lambda_{\rm \!_{H}} + \beta_{\rm \!_{H}})
     n_{\rm \!_{H}} (1-x_{\rm p}) \right)},\nonumber
\end{eqnarray}
\begin{eqnarray}
\label{eq:xHe_evol}
{dx_{\rm \!_{He}}\over dz} =&
   \left(x_{\rm \!_{He}}x_{\rm e} n_{\rm \!_{H}} \alpha_{\rm \!_{He}}
   - \beta_{\rm \!_{He}} (f_{\rm \!_{He}}-x_{\rm \!_{He}})
   {\rm e}^{-h_{\!_{P}} \nu_{\!_{He2^1s}}/k_{\!_{B}} \tk}\right)\\
   &\times {\left(1 + K_{\rm \!_{He}} \Lambda_{\rm \!_{He}} n_{\rm \!_{H}}
  (f_{\rm \!_{He}}-x_{\rm \!_{He}}){\rm e}^{-h_{\!_{P}} \nu_{\!_{ps}}/k_{\!_{B}} \tk})\right)
  \over H(z)(1+z)\left(1+K_{\rm \!_{He}}
  (\Lambda_{\rm \!_{He}} + \beta_{\rm \!_{He}}) n_{\rm \!_{H}} (f_{\rm \!_{He}}-x_{\rm \!_{He}})
  {\rm e}^{-h_{\!_{P}} \nu_{\!_{ps}}/k_{\!_{B}} \tk}\right)}.\nonumber
\end{eqnarray}
The total electron fraction ($x_{\rm e}$) is subsequently defined to be 
\begin{eqnarray}
\label{eq:xe_evol}
    x_{\rm e} = x_{\rm p} + x_{\rm \!_{He}},
\end{eqnarray}
where $x_{\rm p}\equiv \frac{n_{\rm p}}{n_{\!_{\rm H}}}$ and $x_{\rm \!_{He}}\equiv \frac{n_{\rm \!_{He}}}{n_{\!_{\rm H}}}$ are the proton fraction and singly ionized helium fraction, respectively, with $n$ referring the number density (see appendix~\ref{app:ionization_fraction} for further details.). Finally, the evolution of the gas kinetic temperature ($T_k$) of the IGM is expressed as~\cite{PhysRevD.74.103502}
\be\label{eq:evolution_of_gas_temperature}
\frac{d\tk}{dz} = \frac{2\tk}{1+z} - \frac{2}{3H(z)(1+z)}\sum_i \frac{\epsilon_i}{k_{\!_{B}} n_{\rm tot}},
\ee
where the first term represents adiabatic cooling due to cosmic expansion, and each subsequent $\epsilon_i$ term accounts for the energy injection/extraction rate per unit volume corresponding to the $i$\textsuperscript{th} process. In the baseline $\Lambda$CDM scenario,  two effects may reflect on the IGM temperature on top of adiabatic cooling:
\begin{enumerate}[i)]
\item {\it Compton scattering between electrons and photons:} The rate of energy transfer per unit volume due to Compton effect can be expressed as~\cite{Bharadwaj_2004} 
\be \label{eq:epsComp}
\epsilon_{\rm comp} = \frac{3}{2}n k_{\!_{B}} \frac{x_{\rm e}}{1 + x_{\rm e} + f_{\!_{\rm He}}}\frac{8 \sigma_{\!_{T}} u_{\gamma}}{3m_e c}\left(\tgamma - \tk\right)\:,
\ee
with $\sigma_{\!_{T}}$ and $u_{\gamma}$ respectively denoting the Thomson cross-section and the energy density of background photons.

\item {\it X-ray heating:} The X-ray photons produced from galaxies and clusters after star formation can heat the IGM at comparatively lower redshifts, for which the energy injection rate per unit volume is modelled as ~\cite{PhysRevD.74.103502}
\be \label{eq:epsX}
\epsilon_X = 3.4\times 10^{33} f_{\rm heat} f_{\!_{\rm X}} \frac{\sfrd (z)}{M_{\rm sun}\, \rm yr^{-1}\, \rm Mpc^{-3}} \,\rm J\,\rm s^{-1}\,\rm Mpc^{-3}\:,
\ee
with $M_{\rm sun}$ being the mass of the Sun. Additionally, $f_{\rm heat}$ and $f_{\!_{\rm X}}$ are respectively the X-ray heating fraction and the normalization factor accounting for the difference between local and high redshift observations, both of which we fix at $0.2$ based on Refs.~\cite{Chatterjee:2019jts, 2003MNRAS.340..210G, PhysRevD.74.103502}\footnote{In the absence of strongly constraining datasets, the astrophysical parameters have been fixed at these benchmark order-of-magnitude values which are well-motivated from simulation-based studies. While $f_X$ in particular remains quite poorly constrained \cite{HERA:2021noe} and is expected to be degenerate with magnetic parameters as far as IGM heating effects are concerned, our results in this work correspond to these particular representative values of the astrophysical parameters. A detailed exploration of the full allowed range of the astrophysical parameters and the resulting degeneracy falls beyond the scope of this work.}. 
\end{enumerate}

Upon the inclusion of PMFs however, there can be non-trivial impact on the matter clustering and the gas kinetic temperature. These non-trivial impacts strongly affect the star formation rate, ionization history and the global 21-cm signal, whose modelling is discussed in the following section.

\subsection{Signatures of PMF on the global 21-cm signal}
\label{subsec:mag21}
The effects of a non-helical stochastic PMF on the global 21-cm signal is expected to arise through two simultaneous pathways. Firstly, the PMF is capable of directly heating the IGM by transferring energy from the magnetic sector to the kinetic sector through the processes of ambipolar diffusion (AD) and decaying magnetohydrodynamic turbulence (DT). Secondly, the PMF can significantly enhance matter clustering at small scales ($k/h\gtrsim1$ Mpc$^{-1}$) by sourcing density perturbations via Lorentz force, thereby resulting in a robust signature on SFRD, which, in turn, strongly affects the $\xalpha$. In what follows, we briefly discuss these two phenomena. We present our analysis in terms of the redshift-evolving smoothed magnetic field amplitude $B(z)$, which may be formally related with the magnetic power spectrum $P_B(k)$ as outlined in Appendix~\ref{app:bzandpbk}.
\subsubsection{Impact of PMF on IGM temperature}
\label{subsubsec:magTk}

AD and DT inject energy into the IGM, resulting in an increment of the gas temperature $T_k$. The energy injection rate per unit volume due to these two effects can be expressed as~\cite{Bera:2020jsg,heating_3,heating_5,Pinto:2008zn}
\be
\epsilon_{\rm ambi}(z)&\approx&\frac{(1-x_{\rm e})}{\gamma\, x_{\rm e}\, (M_{\!_{\rm H}} n_{\!_{b}})^2}\ \frac{E_B^2\,f_{\!_{L}}(n_{\!_{B}}+3)}{L_d^2}\,\label{eq:gamma_ambi},\\
\epsilon_{\rm turb}(z)&=&\frac{1.5\ m\ \left[\ln(1+t_i/t_d)\right]^m}{\left[\ln(1+t_i/t_d)+1.5\ln\{(1+z_i)/(1+z)\}\right]^{m+1}}H(z)\,E_B\,.\label{eq:gamma_turb}
\ee
Here, $\gamma\approx1.94\times10^{11}(T_{k}/K)^{0.375}$ m$^3$kg$^{-1}$s$^{-1}$ is the coupling coefficient between ionized and neutral baryon components, $E_B=|B|^2/8\pi$ is the energy density of the magnetic field (written in Gaussian units), $f_{\!_{L}}(x)\equiv0.8313\left(1-1.02\times10^{-2}x\right)x^{1.105}$, $m\equiv2(n_{\!_{B}}+3)/(n_{\!_{B}}+5)$, $t_i$ and $t_d$ are respectively the physical time scales at which magnetic turbulence becomes dominant and over which it decays and are interrelated as $t_i/t_d\approx14.8(B_0/\textrm{nG})^{-1}(k_D/\textrm{Mpc}^{-1})^{-1}$, and $L_d$ is the magnetic damping length scale defined as $L_d^{-1}=(1+z)k_D$ \cite{Bera:2020jsg}. The smallest relevant value of the damping wavenumber $k_D$ used in these expressions depends on the photon diffusion scale at decoupling, and can be estimated from \cite{Kunze:2013uja} for best fit values of the background cosmological parameters as
\be\label{eq:kD}
k_D(z)\:\approx\:\alpha(z) k_\gamma\quad;\quad\alpha=\left(\dfrac{B(z)}{\sqrt{16\pi\rho_\gamma(z)/3}}\right)^{-1}\:,
\ee
where $k_\gamma$ represents the photon diffusion scale at decoupling. Thus, the magnetic damping wavenumber at decoupling is a function of the magnetic field strength at decoupling ($B_{\rm CMB}$) and the photon energy density at decoupling ($\rho_{\gamma,\textrm{CMB}}$). If one neglects IGM heating effects or assumes them to be negligibly small compared to Hubble expansion, then the frozen-in magnetic field redshifts as $a^{-2}$ and the photon energy density as $a^{-4}$, which removes any redshift dependence from Eq.~\eqref{eq:kD} and allows one to use their present values instead, \textit{i.e.} $B_0$ and $\rho_{\gamma,0}$. This is, in fact, what has been done in the majority of studies carried out in the past. However, at lower redshifts, IGM heating via ambipolar diffusion and decaying turbulence are not necessarily sub-dominant effects compared to Hubble evolution of the magnetic field, and the relative error in estimating the present day $B_0$ from $B_{\textrm{CMB}}$ through the simple relation $B_{\textrm{CMB}}=B_0(1+z_{\rm CMB})^2$ increases for smaller values of the magnetic field strength. Thus, in our analysis, we use $B_{\rm CMB}$ directly as the input parameter in Eq.~\eqref{eq:kD} to estimate $k_D$, which obviates the need to assume any redshift dependence of $B(z)$ \textit{a priori}. For the photon energy density, of course, one needs to use $\rho_{\gamma,\textrm{CMB}}=\rho_{\gamma,0}(1+z_{\rm CMB})^4$. 

Furthermore, Eq.~\eqref{eq:gamma_turb} contains the ratio $t_i/t_d$, which is estimated conventionally in the literature as a function of $B_0$ \cite{Minoda:2018gxj,Bera:2020jsg}. This might seem problematic for an initial value problem such as ours, where we can only obtain $B_0$ by supplying an initial $B_{\rm CMB}$ as input and solving the coupled first order system. Upon closer inspection though, it becomes clear that the dependence of $\epsilon_{\rm turb}(z)$ on $t_i/t_d$ is firstly logarithmic, and secondly further suppressed by an overall factor of $m$ which is much smaller than unity if $n_{\!_{B}}\to-3$, \textit{i.e.} the magnetic spectrum is nearly scale-invariant. These two observations allow us to approximate the value of $B_0$ appearing in $t_i/t_d$ as $B_0\approx B_{\rm CMB}(1+z_{\rm CMB})^{-2}$. Although this relation is not strictly accurate in presence of heating effects as already explained, the error incurred by this approximation in the specific case of Eq.~\eqref{eq:gamma_turb} is negligible in view of the aforementioned factors. Throughout our analysis, we also restrict our attention to a nearly scale-invariant PMF, which further justifies this approximation in Eq.~\eqref{eq:gamma_turb}. We reiterate that this is \textit{not} to be identified/confused with the true value of the present day magnetic field strength, which can only be obtained by solving the full system with appropriate initial conditions. That being said, however little the error might be in the present scenario, we intend to revisit this issue in future and explore some better theoretical modelling of the DT effect within the initial value framework. 

Finally, taking both Eqs.~\eqref{eq:gamma_ambi} and \eqref{eq:gamma_turb} into account, the evolution of the magnetic field strength itself is governed by the equation
\be\label{eq:Eb}
\frac{dE_B}{dz}=4\,\left(\frac{E_B}{1+z}\right)+\frac{1}{(1+z)\,H(z)}\,&\left[\epsilon_{\rm ambi}(z)+\,\epsilon_{\rm turb}(z)\,\right]\:,
\ee
which succinctly captures the rate of energy transfer from the magnetic sector to the IGM. \footnote{The DT equation derived in \cite{Sethi:2004pe} assumes only background Hubble expansion along with dissipation through DT, and does not incorporate any AD effect. For our purposes, we have adapted it as \eqref{eq:gamma_turb}, and to incorporate both the effect of DT and the additional dissipation due to AD, we use \eqref{eq:Eb} as an approximate solution that combines both effects in addition to Hubble expansion. This should not result in a significant loss of accuracy, as AD and DT effects dominate over different redshift epochs. We intend to revisit the issue of simultaneous AD+DT modeling in a more theoretically accurate and self-consistent way in a future work.}

\subsubsection{Impact of PMF on star formation} 
\label{subsubsec:magsfrd}
\begin{figure*}[!ht]
    \centering
    \begin{subfigure}{.49\textwidth}
    \includegraphics[width=\textwidth]{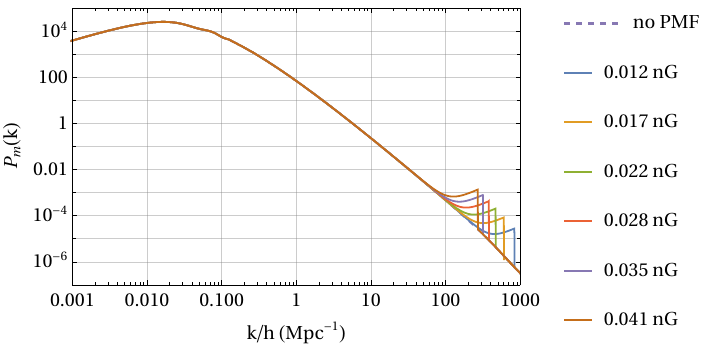}
    \caption{}
    \end{subfigure}
    \begin{subfigure}{.49\textwidth}
    \includegraphics[width=\textwidth]{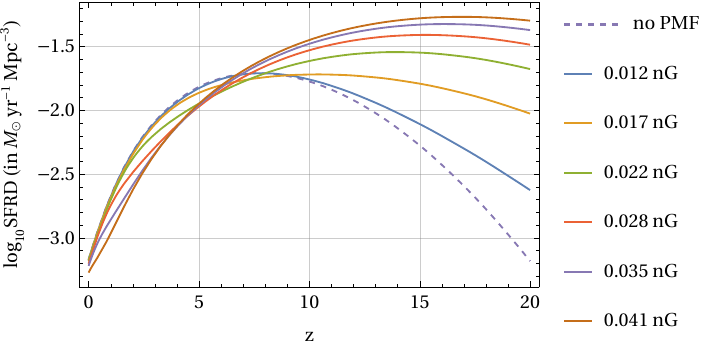}
    \caption{}
    \end{subfigure}
    \begin{subfigure}{.49\textwidth}
    \includegraphics[width=\textwidth]{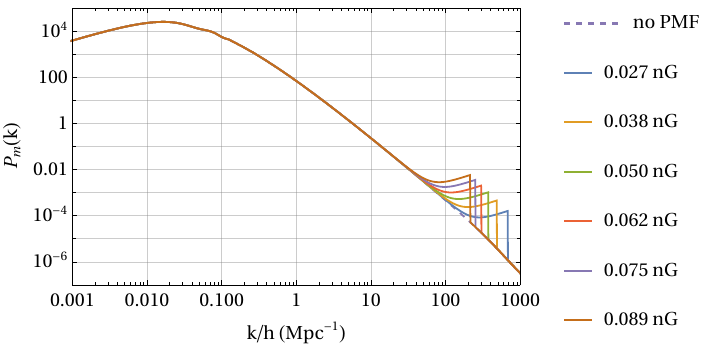}
    \caption{}
    \label{fig:mPk_subfig}
    \end{subfigure}
    \begin{subfigure}{.49\textwidth}
    \includegraphics[width=\textwidth]{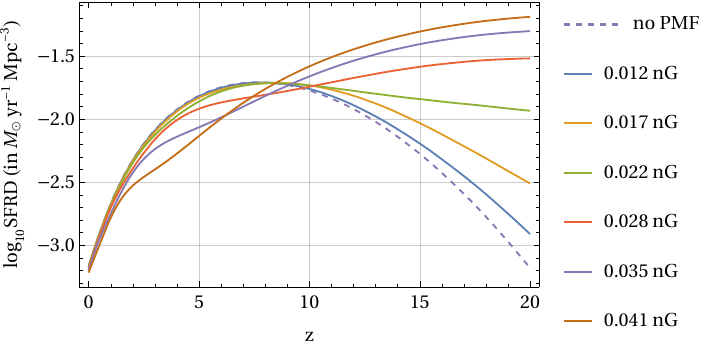}
    \caption{}
    \label{fig:SFRD_subfig}
    \end{subfigure}
    \caption{Impact of primordial magnetic fields on the matter power spectrum and star formation rate density shown for a few representative values of the present day magnetic field strength, with $n_{\!_{B}}=-2.75$ (\textbf{upper panel}) and $n_{\!_{B}}=-2.95$ (\textbf{lower panel}). Both sets of $B_0$ correspond to the same set of $B_{\rm CMB}$ ranging from $7.5\times10^4$ nG to $2.0\times10^5$ nG (in steps of $1.5\times10^4$ nG), thus also reflecting the impact of $n_{\!_{B}}$ on the redshift evolution of $B(z)$ through IGM heating effects.}
    \label{fig:mPk_and_SFRD}
\end{figure*}

The presence of primordial magnetic fields modifies the matter power spectrum on small scales ($k/h\gtrsim1$ Mpc$^{-1}$) through the appearance of an effective Lorentz force source term in the evolution equation of matter density perturbations \cite{PhysRevD.81.043517,PhysRevD.86.043510,2012ApJ...748...27P,Kunze:2013hy,Adi:2023doe}. The modification shows up primarily as a bump on top of the usual adiabatic matter power spectrum, with its position and height governed by two parameters: the present-day magnetic field strength ($B_0$) and the magnetic spectral index ($n_{\!_{B}}$). Up to the linear order in perturbation theory, the dominant correction term on top of the adiabatic matter power spectrum $P_m^{(ad)}(k)$ at present time is given by \cite{Kunze:2022mlr}
\be\label{eq:Pmbf}
P_m^{(B)}(k)=\dfrac{2\pi^2}{k^3}\left(\dfrac{k}{a_0H_0}\right)^4\dfrac{4}{225}(1+z_{eq})^2\left(\dfrac{\Omega_{\gamma,0}}{\Omega_{m,0}}\right)^2\mathcal{P}_{L^{(0)}}(k)\:,
\ee
where $\Omega_{\gamma,0}$ is the present-day photon energy fraction and $z_{eq}$ is the redshift of matter-radiation equality. The dimensionless power spectrum $\mathcal{P}_{L^{(0)}}(k)$ arises due to the Lorentz source term, and can be expressed analytically as
\begin{eqnarray} \label{eq:Plorentz}
    \mathcal{P}_{L^{(0)}}(k) && =\dfrac{9}{\left[\Gamma\left(\frac{n_{\!_{B}}+3}{2}\right)\right]^2}\left(\dfrac{\rho_{B,0}}{\rho_{\gamma,0}}\right)^2\left(\dfrac{k}{k_D}\right)^{2(n_{\!_{B}}+3)}e^{-\left(\frac{k}{k_D}\right)^2} \nonumber\\
    && \times\int\limits_{0}^{\infty}dh h^{n_{\!_{B}}+2}e^{-2\left(\frac{k}{k_D}\right)^2h^2}\int\limits_{-1}^{+1}dx e^{2\left(\frac{k}{k_D}\right)^2hx}(1-2hx+h^2)^{\frac{n_{\!_{B}}-2}{2}} \nonumber\\
    && \times\left[1+2h^2+(1-4h^2)x^2-4hx^3+4h^2x^4\right]\:,
\end{eqnarray}
where $(\rho_{B,0}/\rho_{\gamma,0})\approx9.545\times10^{-8}(B_0/\textrm{nG})^2$ is the ratio of the present-day magnetic energy density and the photon energy density. While computing the magnetic correction, we make use of the magnetic Jeans cut-off scale given by \cite{Sethi:2004pe,Kunze:2013uja}
\be\label{eq:kj}
k_J=\left[14.8\left(\dfrac{\Omega_m}{0.3}\right)^{0.5}\left(\dfrac{h}{0.7}\right)\left(\dfrac{B_0}{1\:\textrm{nG}}\right)^{-1}\right]^{\frac{2}{n_{\!_{B}}+5}}\:\textrm{Mpc}^{-1}\:,
\ee
which appears as a hard cut-off to the magnetically induced matter power spectrum in our calculations. The total matter power spectrum, $P_m^{(\rm tot)}(k)$, is then given by $P_m^{(\rm tot)}(k)=P_m^{(\rm ad)}(k)+P_m^{(B)}(k)$. In the linear approximation, the matter power spectrum at any earlier epoch with $z>0$ is given by $P_m^{(\rm tot)}(k;z)\approx D(z)^2P_m^{(\rm tot)}(k)$, where $D(z)=(1+z)^{-1}$ is the normal growth factor. \footnote{Strictly speaking, the magnetically corrected growth factor should deviate from $D(z)$ at higher redshifts, but may be approximated with $D(z)$ up to $\sim93\%$ accuracy for $z\lesssim30$ \cite{structure_4}, which is the region of our interest.} Based on the total matter power spectrum, the SFRD $\sfrd (z)$ can then be calculated according to Eq.~\eqref{eq:sfrd}. The magnetically modified redshift-dependence of the SFRD has been shown for a few benchmark values of the PMF parameters in Fig. \ref{fig:mPk_and_SFRD}. \footnote{In Ref.~\cite{structure_2}, the effect of magnetic fields on the filtering mass scale for halo collapse has been investigated, where the authors report that $M_{\rm min}$ may increase due to IGM heating and lead to early suppression of the SFRD. However, their study is based on an extremely blue Batchelor spectrum with $n_{\!_{B}}=3$, which leads to significant enhancement of heating effects (\textit{viz.} Eqs. \eqref{eq:gamma_ambi} \& \eqref{eq:gamma_turb}) compared to $\nb\to-3$. Secondly, appreciable change in the filtering scale is seen only for $B_0\gtrsim0.03$ nG even in their scenario, which exceeds the typical upper bounds from EDGES as we find in Sec. \ref{sec:edgesbound}. These observations lead us to believe that the effect of the nearly scale-invariant PMF on the filtering mass should be negligible for our present purpose.}

\section{Bounds on PMFs from EDGES data}
\label{sec:edgesbound}

Accounting for both IGM heating effects (AD and DT) and the modified star formation history in presence of non-helical PMFs, we now proceed to estimate the admissible bounds on the PMF parameters based on the global 21-cm signal observed by the EDGES collaboration~\cite{Bowman:2018yin}. The system of first order differential equations to be solved with appropriate initial conditions (defined at decoupling) comprises of Eqs. \eqref{eq:xp_evol}, \eqref{eq:xHe_evol}, \eqref{eq:evolution_of_gas_temperature}, and \eqref{eq:Eb}. 

As already mentioned, most of the recent analytical studies conducted  in this direction either focus solely on IGM heating and do not account for the effect of the PMF on the SFRD \cite{Minoda:2018gxj,Bhatt:2019lwt,Natwariya:2020ksr,Natwariya:2020mhe,Bera:2020jsg}, or consider only the effects of the modified star formation history on the global 21-cm signal \cite{Cruz:2023rmo}, thus missing out on key individual pieces of information which are crucial when it comes to constraining PMFs via the available 21-cm window. In this work, we have attempted to address this issue by integrating both phenomena within a common analytical framework. Such an approach has its own challenges, to highlight which we first briefly discuss our adopted pipeline. Calculating the SFRD using Eq.~\eqref{eq:sfrd} requires knowledge of the present-day magnetic field strength $B_0$, whose value, in turn, depends crucially on AD and DT effects in addition to the usual cosmic expansion of the background. On the other hand, modification of the standard star formation history via small-scale enhancement of matter power does not significantly affect the magnetic field strength in itself \cite{Cruz:2023rmo}. Thus, we first solve our system of equations corresponding to a standard SFRD history without the magnetic correction, with the aim of obtaining $B(z)$ and estimating $B_0$ for a given initial value of $B_{\rm CMB}$. This step is important because $B_{\rm CMB}\neq B_0(1+z_{\rm CMB})^2$ in presence of AD and DT unlike what has been assumed \textit{a priori} in a few earlier works, and the deviation from this relation can be significant for the typical magnetic field strengths we are interested in.

Moreover, the AD effect, which dominates at smaller redshifts, is suppressed quickly once reionization sets in and $x_{\rm e}$ starts to increase (\textit{viz.} Eq. \eqref{eq:gamma_ambi}). An accurate description of the redshift evolution of $B(z)$ over the entire redshift range $0<z<1088$ thus depends also on the details of the reionization model, which is not unambiguous in light of present data. This issue has not been adequately addressed in earlier analytical studies, where $B_0$ and $B_{\rm CMB}$ have been related via the simple scale factor relation even in presence of AD and DT effects. In the present analysis, we attempt to take this issue into account by conservatively assuming $x_{\rm e}$ to have reached $\mathcal{O}(1\%)$ by $z\sim12$~\cite{2012RPPh...75h6901P}, and regulating the AD heating term with a resulting sharp $\textrm{tanh}$-cutoff for $z<12$. We have further confirmed that varying the AD cut-off redshift within the range $z\sim8-15$ does not significantly change the resulting value of $B_0$, which remains in the same order-of-magnitude ballpark. Having said that, we must admit that more realistic analytical modelling  and/or case studies may be performed in the light of different viable reionization profiles, which fall beyond the scope of this paper. 

\begin{figure*}[!ht]
    \centering
    \begin{subfigure}{.45\textwidth}
    \includegraphics[width=\textwidth]{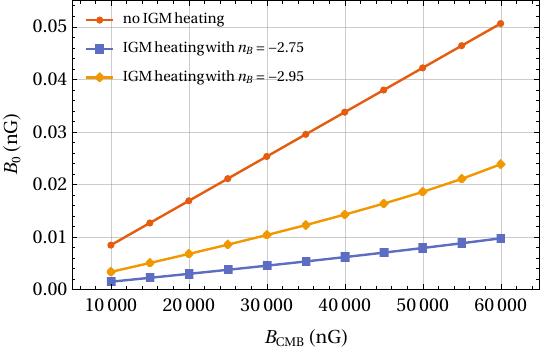}
    \caption{Absolute differences}
    \end{subfigure}
    \begin{subfigure}{.45\textwidth}
    \includegraphics[width=\textwidth]{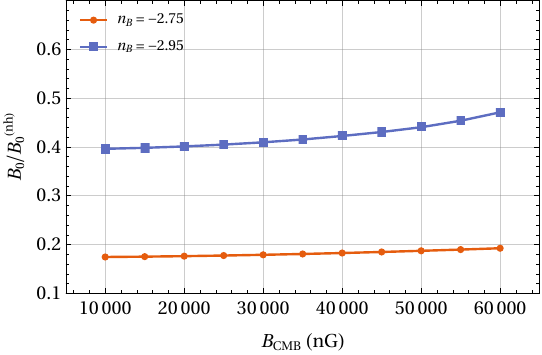}
    \caption{Relative suppression ratio}
    \end{subfigure}
    \caption{ Difference between the values of the present day magnetic field strength ($B_0$) in presence and absence of IGM heating effects via ambipolar diffusion (AD) and decaying turbulence (DT), corresponding to the same set of representative field strengths at recombination ($B_{\rm CMB}$). This shows the significant impact of heating effects on the redshift evolution of weak non-helical PMFs. In the right panel, $B_0^{\rm (nh)}$ denotes the case without IGM heating.}
    \label{fig:b0_bcmb}
\end{figure*}

With these details in view, the non-trivial impact of IGM heating on the redshift evolution of $B(z)$ is presented in Fig.~\ref{fig:b0_bcmb}, where the difference with the standard scale factor evolution $B(z)=B_0(1+z)^2$ has been explicitly shown for two representative values of the magnetic spectral index. As energy is transferred from the magnetic sector to the kinetic sector of the IGM, the magnetic field dilutes away quicker than in the standard scenario. The suppression related to heating effects is found to be more pronounced for weaker field strengths as well as for sharper scale-dependence of the PMF power spectrum. 

In the final step, we use the obtained $B_0$ to compute $P_m^{(B)}(k)$ using Eq.~\eqref{eq:Pmbf}, and subsequently calculate the modified SFRD as a function of redshift using Eq.~\eqref{eq:sfrd}. Finally, we solve the full system of equations with the modified SFRD in place, and proceed to compute $\deltat(z)$ to compare with the observed global 21-cm dip reported by EDGES~\cite{Bowman:2018yin}. A non-helical PMF alone, due to its heating effect, makes the global 21-cm dip shallower and leads to a worsened discrepancy with observations compared to vanilla $\Lambda$CDM. However, an excess extra-galactic radio background~\cite{Subrahmanyan:2013eqa}, supported by ARCADE-2~\cite{fixsen2011arcade} and LWA1~\cite{2012JAI.....150004T}, provides an effective cooling effect and is known to be capable of extending the $\deltat$ dip downward inside the observed bounds from EDGES~\cite{Feng:2018rje}. This excess radio background profile can be modelled as~\cite{fixsen2011arcade}
\be\label{arcade}
T(\nu) = T_{\rm CMB} + \xi T_R\left(\frac{\nu}{\nu_0}\right)^{\beta},
\ee
with $T_R = 1.19\pm0.14$ K, $\beta=-2.62\pm0.04$ and $\nu_0 = 1$ GHz being the fitting parameters and $\xi$ being the fraction of excess radiation considered in the process \cite{feng2018enhanced}.

In the present analysis,  we compute the full redshift profiles of $\deltat(z)$ corresponding to two benchmark values of the magnetic spectral index $n_{\!_{B}}=-2.75$ and $n_{\!_{B}}=-2.95$ (which represent a nearly scale-invariant magnetic power spectrum), considering six different representative values of $\xi=\{0\%,\:2\%,\:5\%,\:10\%,\:15\%,\:20\%\}$. In each case, we show the behaviour of the global 21-cm signal for ten representative values of $B_{\rm CMB}$ ranging from $1.0\times10^4$ nG to $6.0\times10^4$ nG (at incremental steps of $5.0\times10^3$ nG) for $n_{\!_{B}}=-2.95$, and from $1.0\times10^4$ nG to $1.0\times10^5$ nG (at incremental steps of $1.0\times10^4$ nG) for $n_{\!_{B}}=-2.75$. For the two different values of $n_{\!_{B}}$, such different sets of $B_{\rm CMB}$ values have been chosen to better depict the effects on $\deltat(z)$ and obtain upper bounds in light of the EDGES data, as the PMF scale-dependence plays a crucial role in the dynamics of the AD and DT effects. Incorporating the heating phenomena and star formation, both sets of $B_{\rm CMB}$ values typically correspond to $\mathcal{O}(10^{-3}-10^{-2})$ nG field strengths for $B_0$. 

\begin{figure*}[!ht]
    \centering
    \begin{subfigure}{.49\textwidth}
    \includegraphics[width=\textwidth]{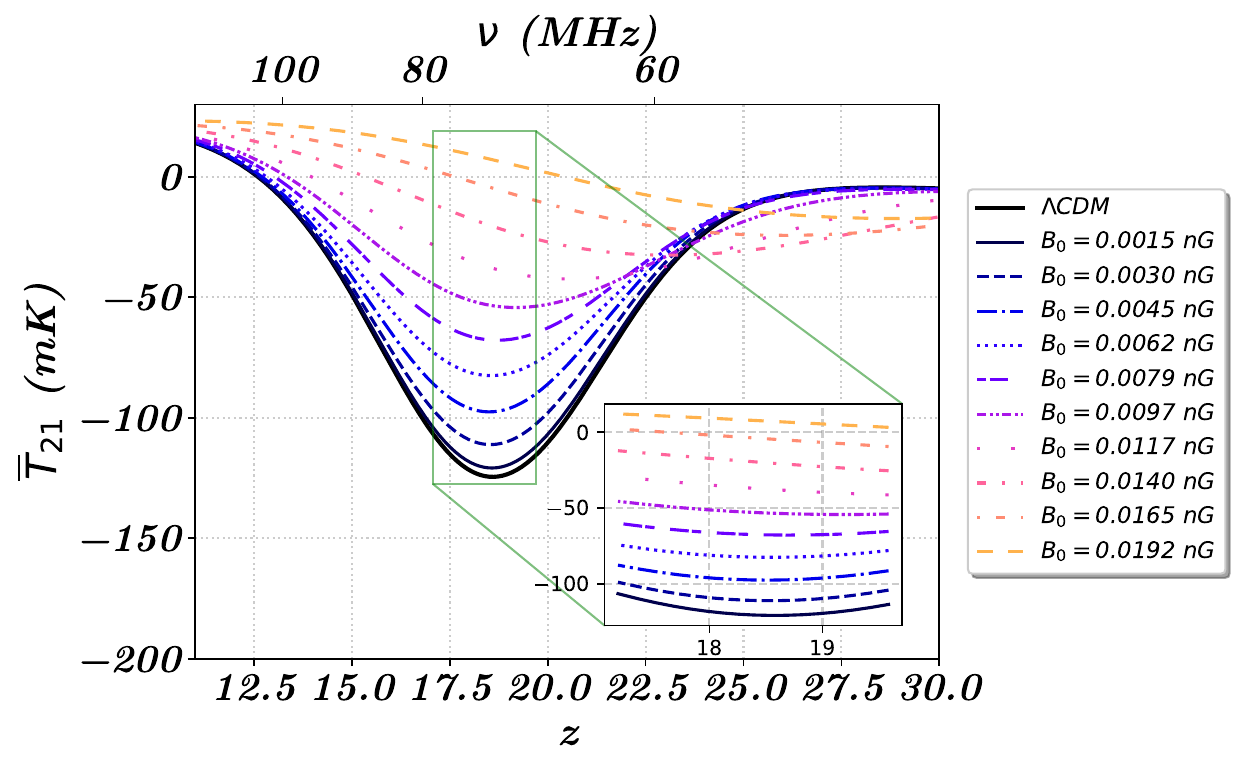}
    \caption{ $\xi=0\%$ (absence of excess radio background)}
    \label{fig:T21_2.75_xi=0}
    \end{subfigure}
    \begin{subfigure}{.49\textwidth}
    \includegraphics[width=\textwidth]{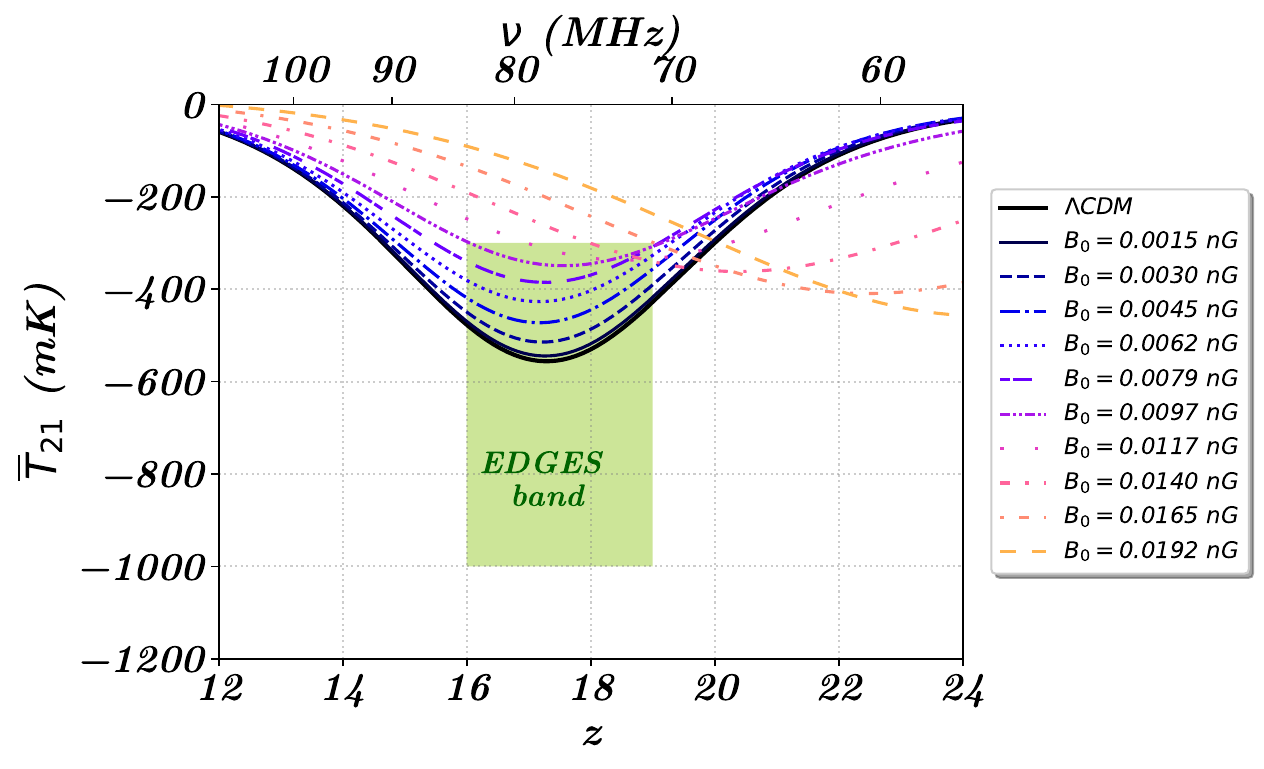}
    \caption{$\xi=2\%$}
    \label{fig:T21_2.75_xi=0.02}   
    \end{subfigure}
    \begin{subfigure}{.49\textwidth}
    \includegraphics[width=\textwidth]{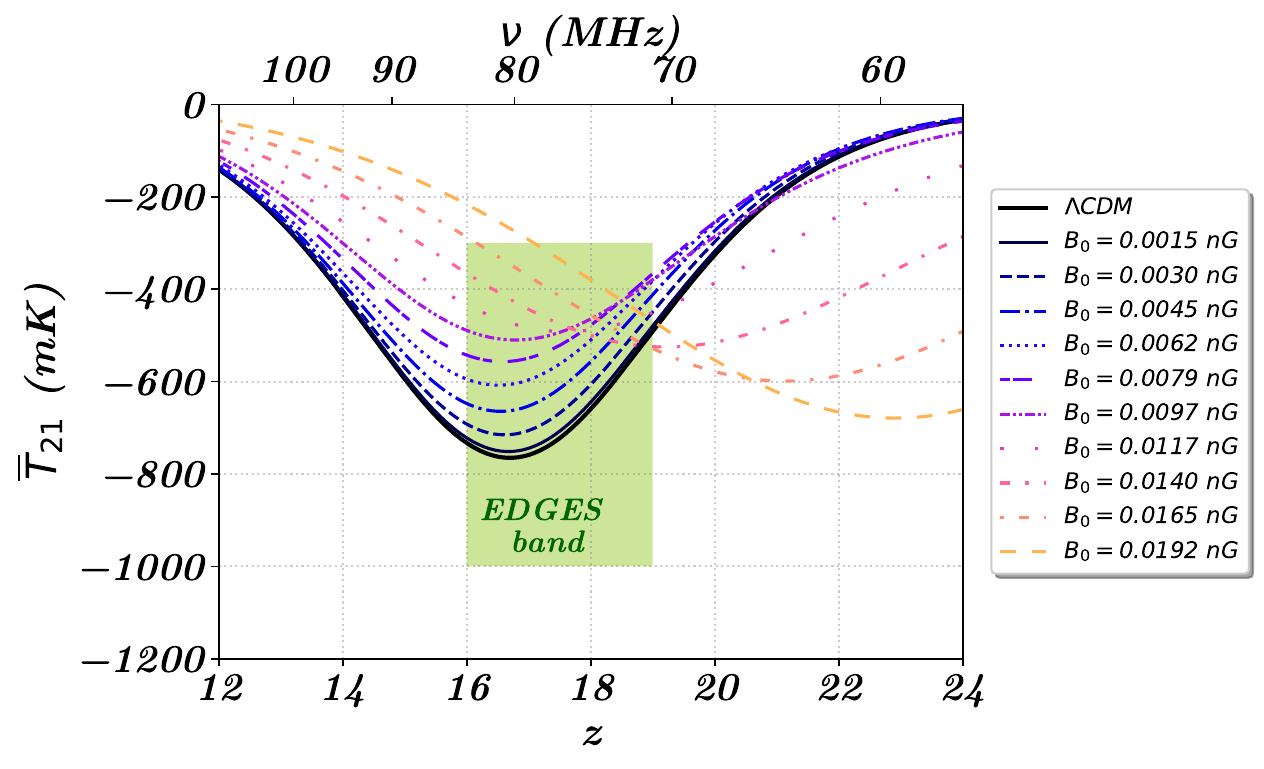}
    \caption{$\xi=5\%$}
    \label{fig:T21_2.75_xi=0.05}   
    \end{subfigure}
    \begin{subfigure}{.49\textwidth}
    \includegraphics[width=\textwidth]{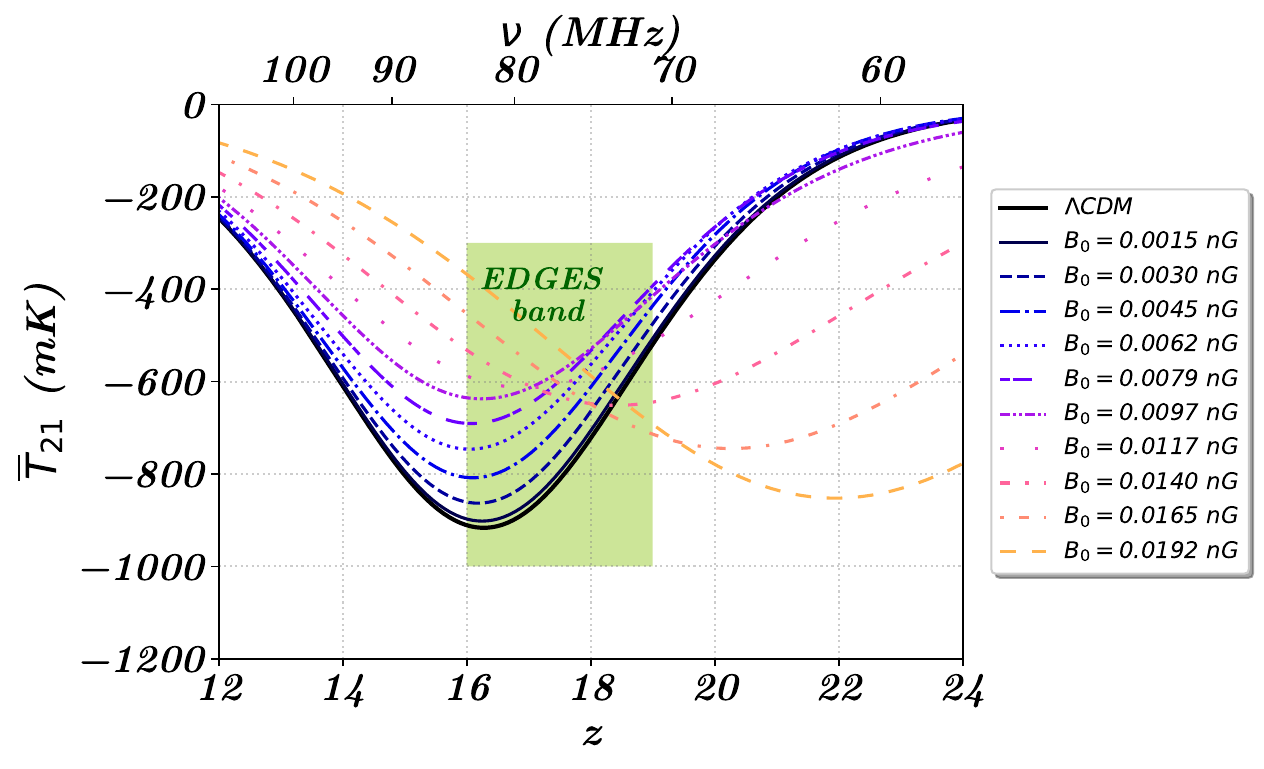}
    \caption{$\xi=10\%$}
    \label{fig:T21_2.75_xi=0.1}   
    \end{subfigure}
    \begin{subfigure}{.49\textwidth}
    \includegraphics[width=\textwidth]{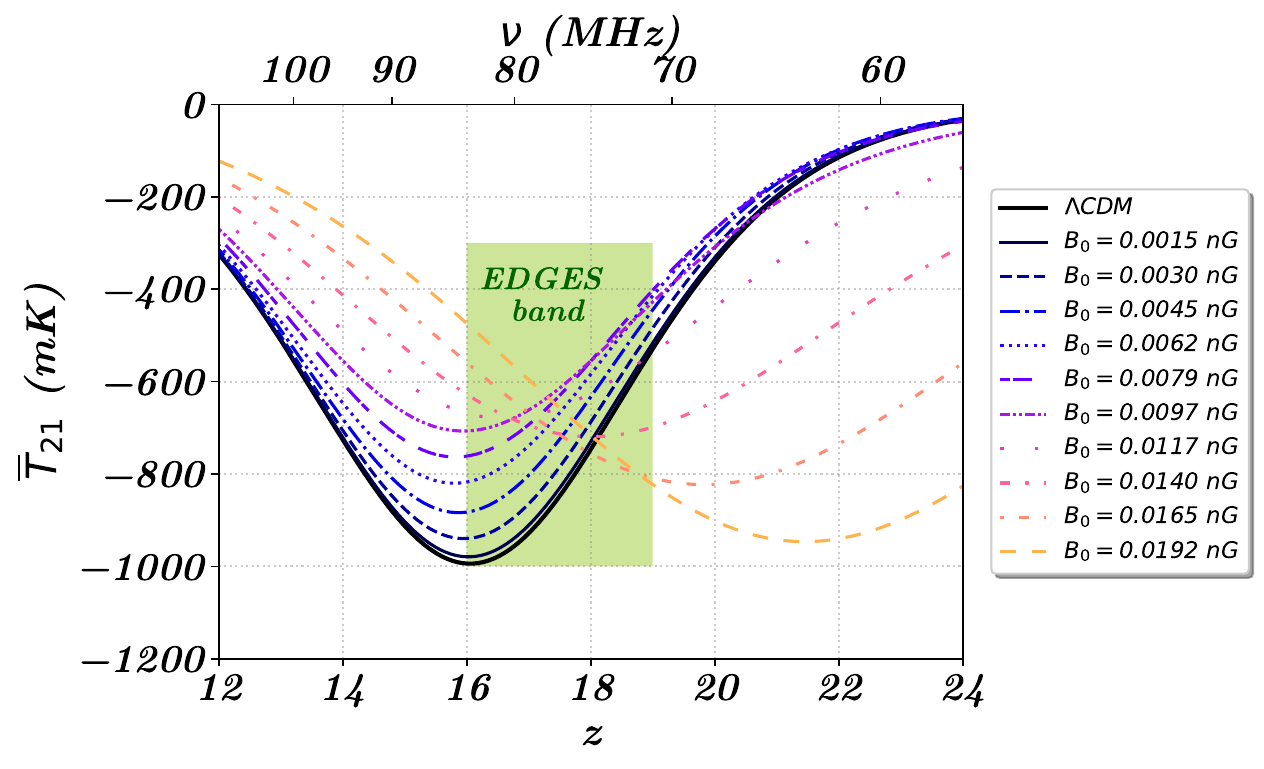}
    \caption{$\xi=15\%$}
    \label{fig:T21_2.75_xi=0.15}   
    \end{subfigure}
    \begin{subfigure}{.49\textwidth}
    \includegraphics[width=\textwidth]{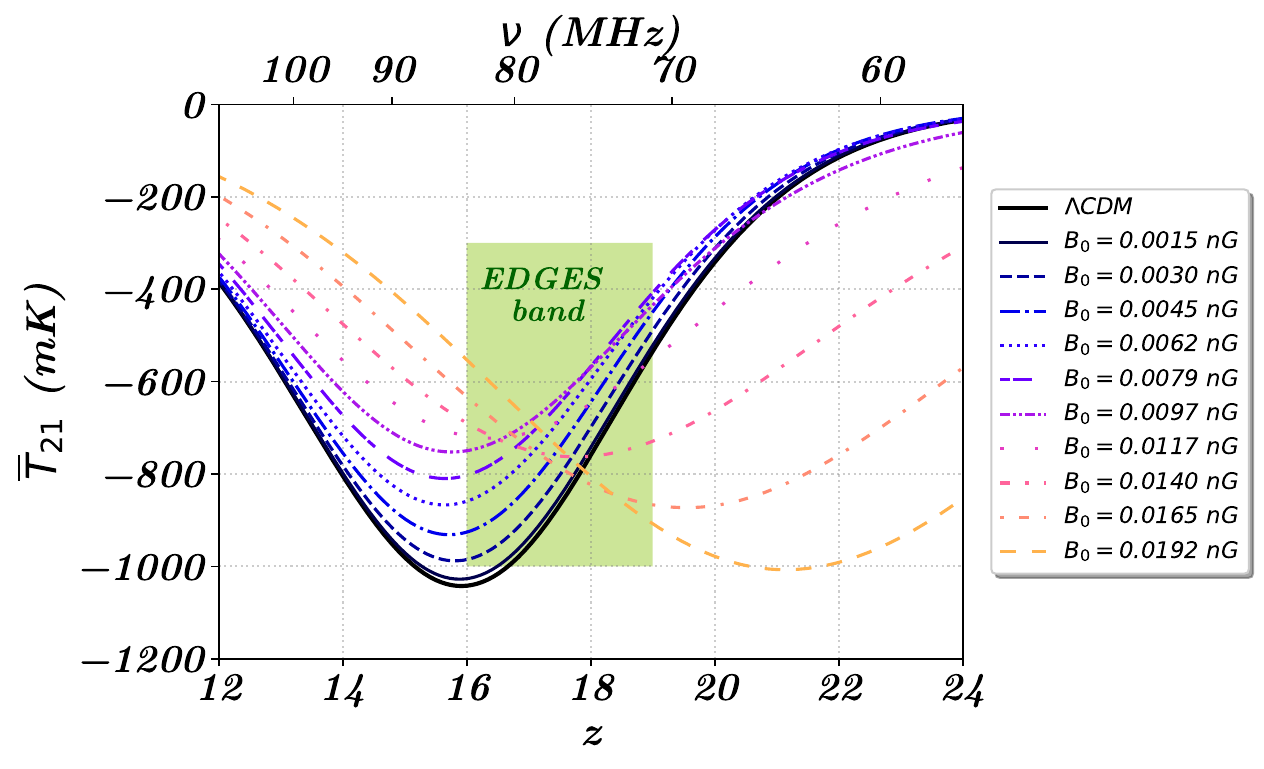}
    \caption{$\xi=20\%$}
    \label{fig:T21_2.75_xi=0.2}   
    \end{subfigure}
    \caption{ Global differential brightness temperature for different values of the ARCADE excess radiation parameter ($\xi$), where $n_{\!_{B}}=-2.75$ has been assumed. The EDGES bounds and $\Lambda$CDM plot have been shown for comparison. The quoted values of $B_0$ correspond to a set of $B_{\rm CMB}$ values ranging from $1.0\times10^4$ nG to $1.0\times10^5$ nG in steps of $1.0\times10^4$ nG.}
    \label{fig:T_21_2.75}
\end{figure*}

\begin{figure*}[!ht]
    \centering
    \begin{subfigure}{.49\textwidth}
    \includegraphics[width=\textwidth]{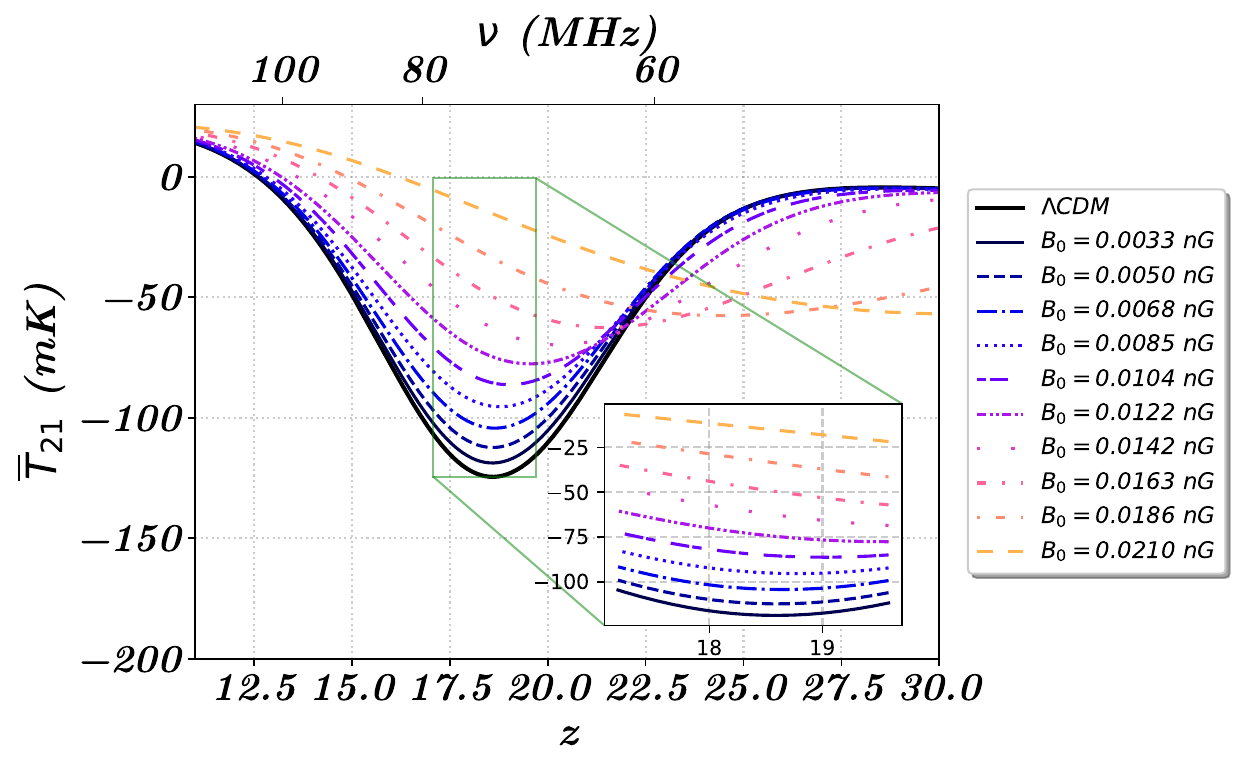}
    \caption{ $\xi=0\%$ (absence of excess radio background)}
    \label{fig:T21_2.95_xi=0}
    \end{subfigure}
    \begin{subfigure}{.49\textwidth}
    \includegraphics[width=\textwidth]{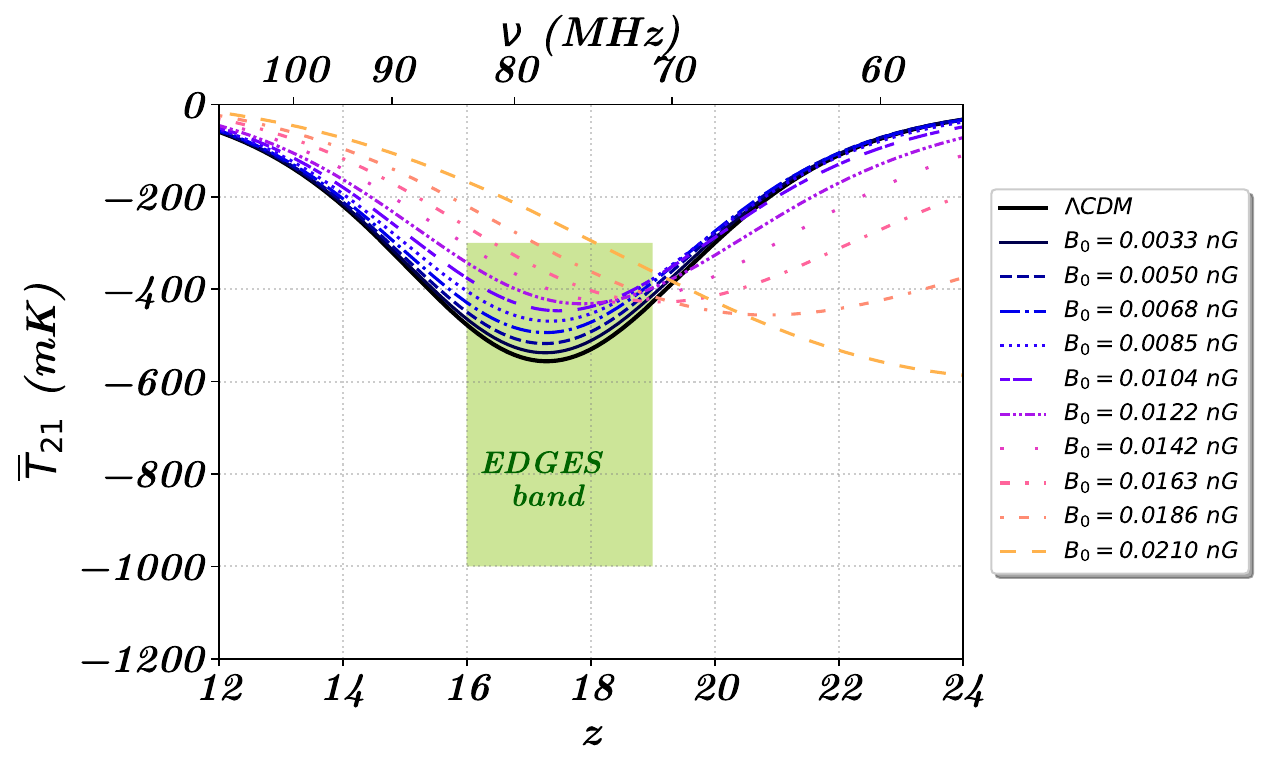}
    \caption{$\xi=2\%$}
    \label{fig:T21_2.95_xi=0.02}   
    \end{subfigure}
    \begin{subfigure}{.49\textwidth}
    \includegraphics[width=\textwidth]{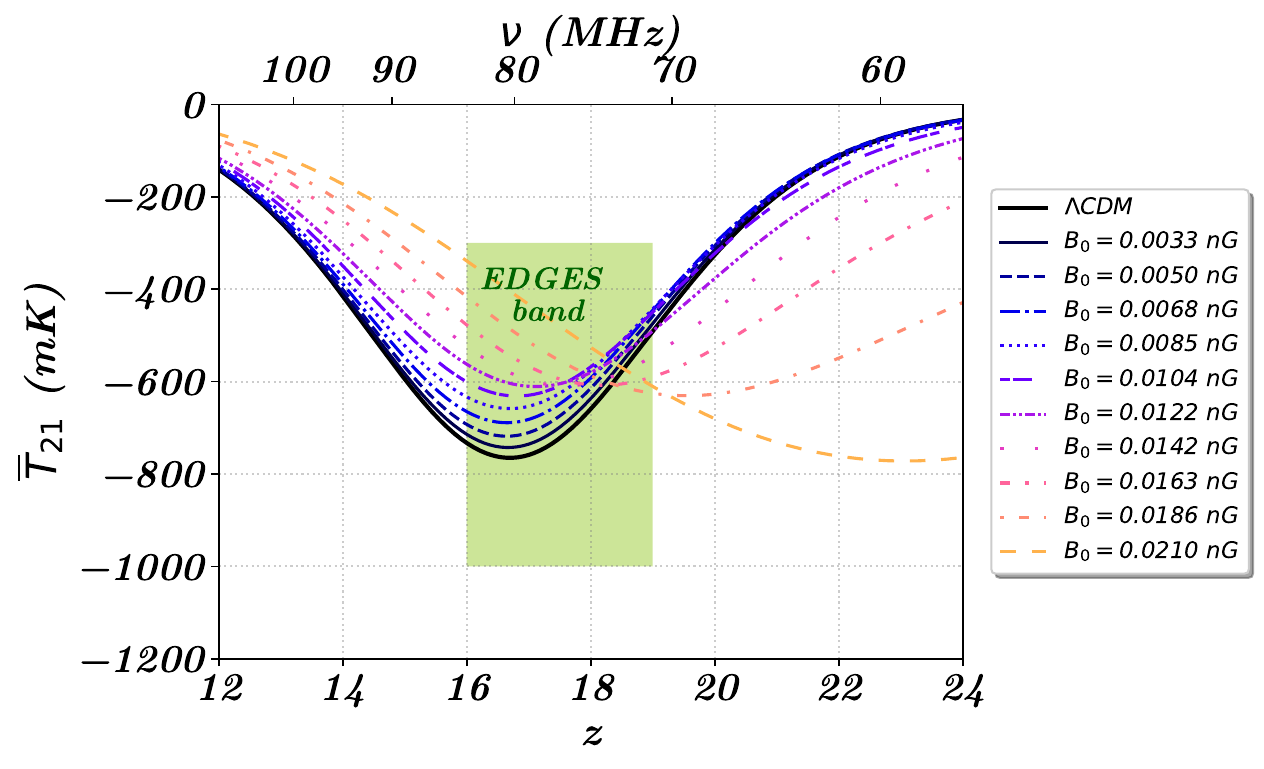}
    \caption{$\xi=5\%$}
    \label{fig:T21_2.95_xi=0.05}   
    \end{subfigure}
    \begin{subfigure}{.49\textwidth}
    \includegraphics[width=\textwidth]{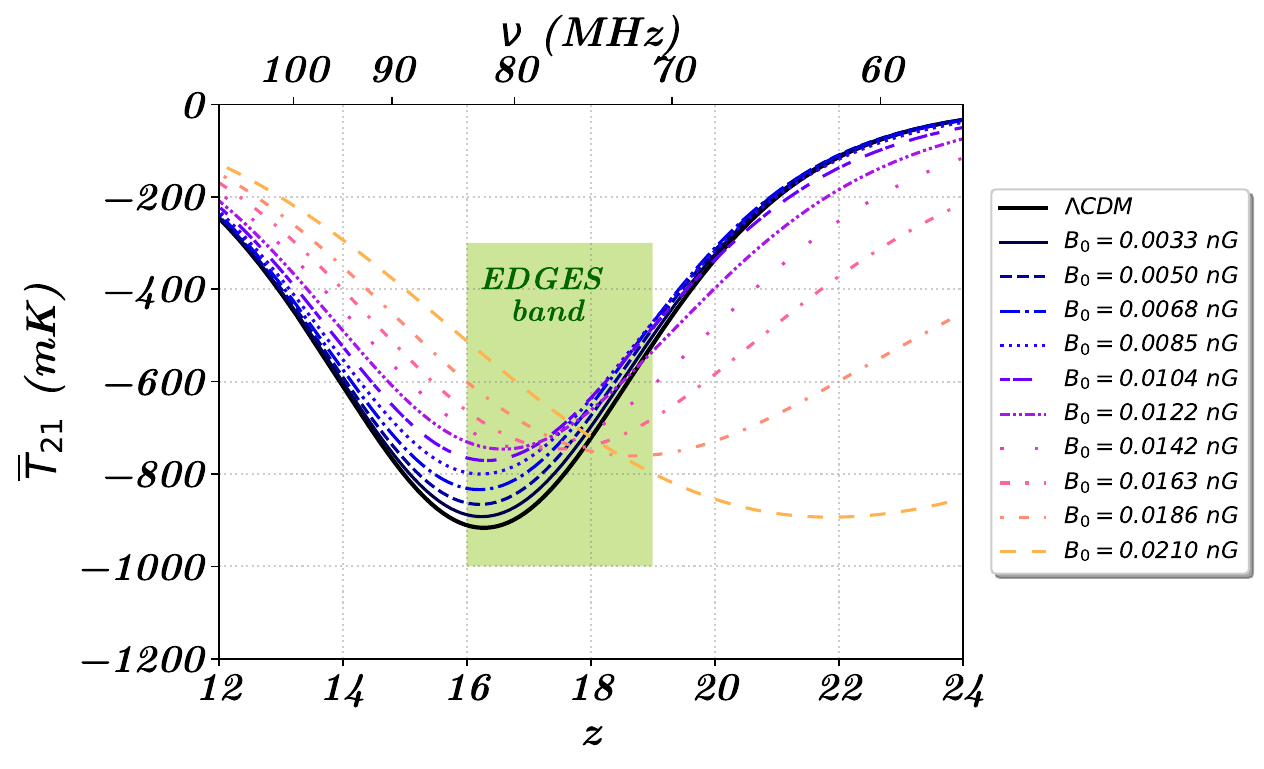}
    \caption{$\xi=10\%$}
    \label{fig:T21_2.95_xi=0.1}   
    \end{subfigure}
    \begin{subfigure}{.49\textwidth}
    \includegraphics[width=\textwidth]{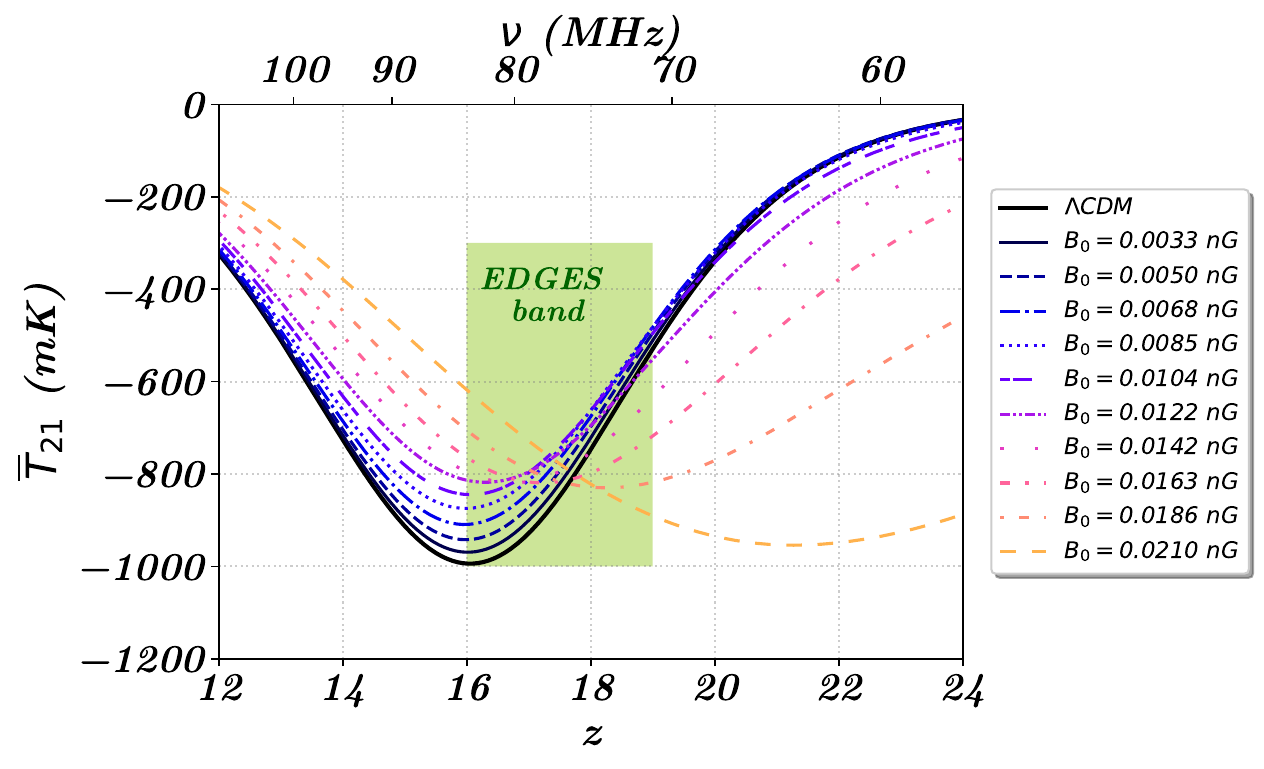}
    \caption{$\xi=15\%$}
    \label{fig:T21_2.95_xi=0.15}   
    \end{subfigure}
    \begin{subfigure}{.49\textwidth}
    \includegraphics[width=\textwidth]{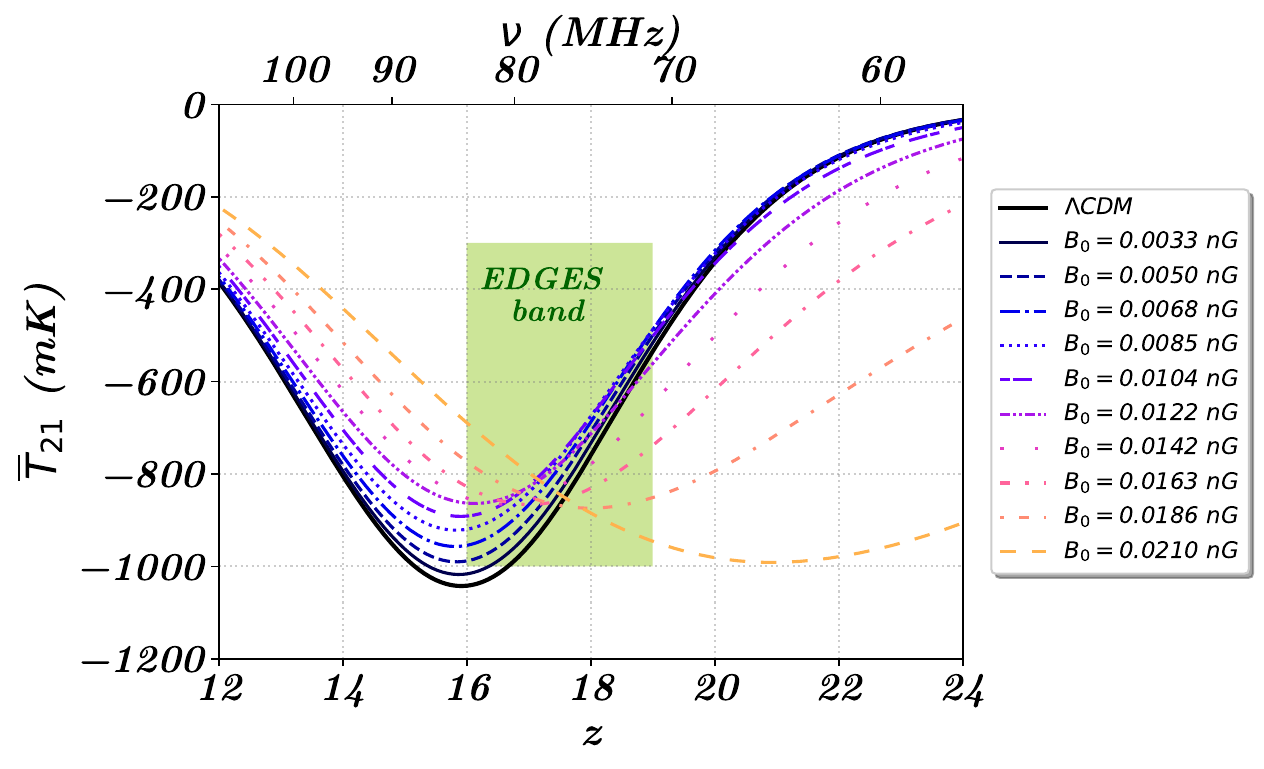}
    \caption{$\xi=20\%$}
    \label{fig:T21_2.95_xi=0.2}   
    \end{subfigure}
    \caption{ Global differential brightness temperature for different values of the ARCADE excess radiation parameter ($\xi$), where $n_{\!_{B}}=-2.95$ has been assumed. The EDGES bounds and $\Lambda$CDM plot have been shown for comparison. The quoted values of $B_0$ correspond to a set of $B_{\rm CMB}$ values ranging from $1.0\times10^4$ nG to $6.0\times10^4$ nG in steps of $5.0\times10^3$ nG.}
    \label{fig:T_21_2.95}
\end{figure*}

The resulting $\deltat(z)$ profiles are shown in Figs. \ref{fig:T_21_2.75} and \ref{fig:T_21_2.95} alongside the observed bounds reported by EDGES \cite{Bowman:2018yin}. Several interesting features immediately stand out. Firstly, while the IGM heating effects dominate for lower values of $B_0$ and gradually raise the 21-cm trough, the effect of enhanced SFRD rapidly becomes prominent as the field strength is increased. The latter induces a sharp lateral shift of the 21-cm trough to progressively earlier redshifts, thereby providing a tighter upper bound on the PMF strength based on the EDGES redshift range $16\lesssim z\lesssim19$ compared to what can be achieved by considering the heating phenomena alone. The upper bound thus obtained varies between $B_0\sim0.015$ nG and $B_0\sim0.018$ nG as $\xi$ is allowed to vary between $2\%$ and $20\%$, as evident from Figs. \ref{fig:T_21_2.75} and \ref{fig:T_21_2.95}. This is nearly two orders of magnitude smaller than the upper bound on $B_0$ in the nearly scale-invariant non-helical regime, which has been reported earlier in \cite{Natwariya:2020ksr} by considering magnetic heating effects alone. For a fixed value of $\nb$, the upper bound increases slightly for higher values of $\xi$, since excess radio background tends to drag the position of the 21-cm trough to slightly lower redshifts. On the other hand, the admissible upper bound on $B_0$ is slightly higher for $\nb=-2.95$ than for $\nb=-2.75$, as seen by comparing Fig. \ref{fig:T_21_2.95} against Fig. \ref{fig:T_21_2.75}. Moreover, an interesting interplay of PMFs with the excess radio background profile used in the modelling of ARCADE data~\cite{fixsen2011arcade} becomes apparent at higher redshifts beyond the EDGES band. In absence of any excess radio background, which is indicated by $\xi=0\%$ in Figs. \ref{fig:T21_2.75_xi=0} and \ref{fig:T21_2.95_xi=0}, the $\deltat$-minima become progressively shallower as they move to higher redshifts with increasing $B_0$, as also noted by Ref.~\cite{Cruz:2023rmo}. Upon the inclusion of the excess radio profile, however, the ``shallowing'' of the minima apparently stops at some critical value of $B_0$, and all subsequent minima at higher redshifts tend to be deeper. This turning point typically occurs beyond the EDGES redshift range which is of interest to us, with the sole exception of the $\xi=20\%$ case which marginally includes the turning point. 

The second interesting feature, which becomes clear upon careful inspection of Figs. \ref{fig:T_21_2.75} and \ref{fig:T_21_2.95}, occurs for the smaller values of $B_0$ depicted therein. As $B_0$ starts to increase and the minima become shallower due to magnetic heating of the IGM, there is a very slight initial leftward shift of the dip to lower redshifts. At first glance, this appears contrary to what one may anticipate, as higher field strengths are expected to enhance early SFRD in our modelling. However, this can be explained on the basis of the redshift-dependence of the magnetically enhanced SFRD, which has been plotted in the right column of Fig. \ref{fig:mPk_and_SFRD} for a few (larger) benchmark values of $B_0$. For a given pair of neighbouring values $B_{0}^{(1)}<B_{0}^{(2)}$, the two corresponding SFRD curves cross each other at some particular redshift ($z_{\rm cross}$), with the SFRD for $B_{0}^{(2)}$ being less than that of $B_{0}^{(1)}$ for $z<z_{\rm cross}$. As visible in Fig. \ref{fig:SFRD_subfig}, this crossing takes place at comparatively higher redshifts for weaker field strengths, and occur earlier compared to the EDGES interval for the weakest values of $B_0$ depicted in Figs. \ref{fig:T_21_2.75} and \ref{fig:T_21_2.95}. Thus, the SFRD slightly decreases in the redshift range $z\lesssim19$ as $B_0$ is increased from the bottom up, resulting in the marginal leftward shift of the 21-cm minima. As $B_0$ subsequently increases past a threshold value, the crossing redshift moves to the left of the EDGES interval, and the SFRD subsequently rises in our region of interest with further increase in $B_0$, producing the pronounced rightward shift of the dip. While this minute effect is hard to glean from the plots (Figs. \ref{fig:T21_2.75_xi=0} and \ref{fig:T21_2.95_xi=0} show zoomed in panels of this region), it is nevertheless significant and plays an important role in determining the correlations among various parameters with respect to their fiducial values, as we investigate in Sec.~\ref{subsec:fisher_results}.

\section{Impact of PMF on the 21-cm power spectrum}
\label{sec:powspec}

Let us now move on to discuss the effects of PMF on the 21-cm power spectrum that would in turn help us in forecasting on the upcoming 21-cm mission SKA-Low.
The two-point function for differential brightness temperature fluctuations is defined as~\cite{Munoz:2015eqa}
\be \label{eq:21cmpowspecdef}
\langle\delta T_{21}(\bold{k};z)\delta T_{21}(\bold{k'};z)\rangle \equiv P_{21}(\bold{k};z)\delta^{(3)}(\bold{k}+\bold{k'})\:,
\ee
where $\delta T_{21}(\bold{k};z)$ is the Fourier transform of the real-space temperature difference $\delta T_{21}(\bold{x};z) \equiv T_{21}(\bold{x};z)-\deltat(z)$. Here $P_{21}(\bold{k};z)$ represents the 21-cm power spectra which can be modelled analytically as~\cite{Munoz:2015eqa}
\be \label{eq:21cmpowspecexpr}
P_{21}(\bold{k};z)=\left[\mathcal{A}(z)+\deltat(z)\mu^2\right]^2P_{\rm HI}(k;z)\:,
\ee
where $P_{\rm HI}(k;z)$ is the power spectrum of the neutral hydrogen density fluctuations which is approximately equal to the baryon power spectrum $P_b(k;z)$ within the redshift range under consideration, $\mu\equiv k_\parallel/k$ is the cosine of the angle between the line of sight component ($k_\parallel$) and the absolute value ($k$). Furthermore, $\mathcal{A}(z)\equiv \frac{d\overline{T}_{21}(z)}{d\delta_b}$~\cite{Munoz:2015eqa,Ali-Haimoud:2013hpa} which can be approximated as $\mathcal{A}(z)\approx 2\overline{T}_{21}(z)$ for $z\lesssim 50$~\cite{Ali-Haimoud:2013hpa}, and $\delta_b$ is the baryon density fluctuation. For observational purposes, one needs to average over all directions to obtain the 1D power spectrum $P_{21}(k;z)$, which translates to an integral of Eq.~\eqref{eq:21cmpowspecexpr} over $\mu$. The 21-cm power spectrum thus effectively traces the baryon distribution as a function of $k$ with a redshift-dependent bias.

Like the total matter power spectrum, $P_b(k;z)$ is similarly enhanced on small scales in presence of PMFs. However, for the typical magnetic field strength which is of interest to us, this enhancement occurs at $k\gtrsim100$ Mpc$^{-1}$, which lies beyond the observable window of next-generation intensity mapping experiments like the Square Kilometre Array (SKA). Hence, limiting our attention to the scales $k\lesssim10$ Mpc$^{-1}$ that are of observational interest, the only effect of PMFs on $P_{21}(k;z)$ arises through $\deltat(z)$. Thus, in our calculations, we make use of the $\deltat(z)$, computed in Sec.~\ref{sec:edgesbound} and the numerically obtained $P_b(k;z)$ from \texttt{CLASS}~\cite{2011arXiv1104.2932L,2011JCAP...07..034B} for each redshift slice. 

\begin{figure*}[!ht]
    \centering
    \begin{subfigure}{.49\textwidth}
    \includegraphics[width=\textwidth]{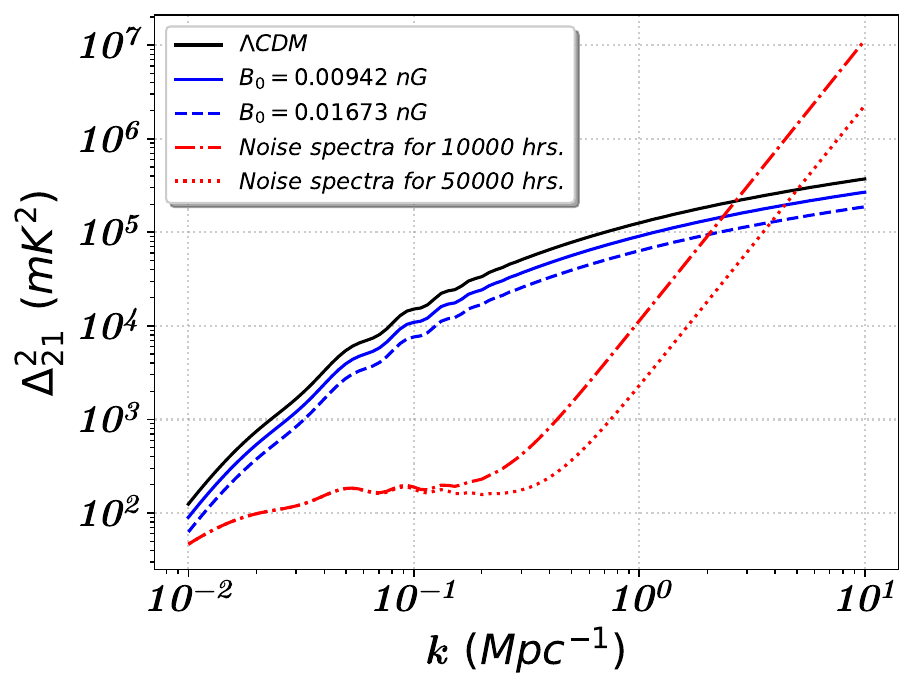}
    \caption{ Momentum profiles wrt $B_0$ at $z=16$}
    \label{fig:P21_noise_zfixed}
    \end{subfigure}
    \begin{subfigure}{.49\textwidth}
    \includegraphics[width=\textwidth]{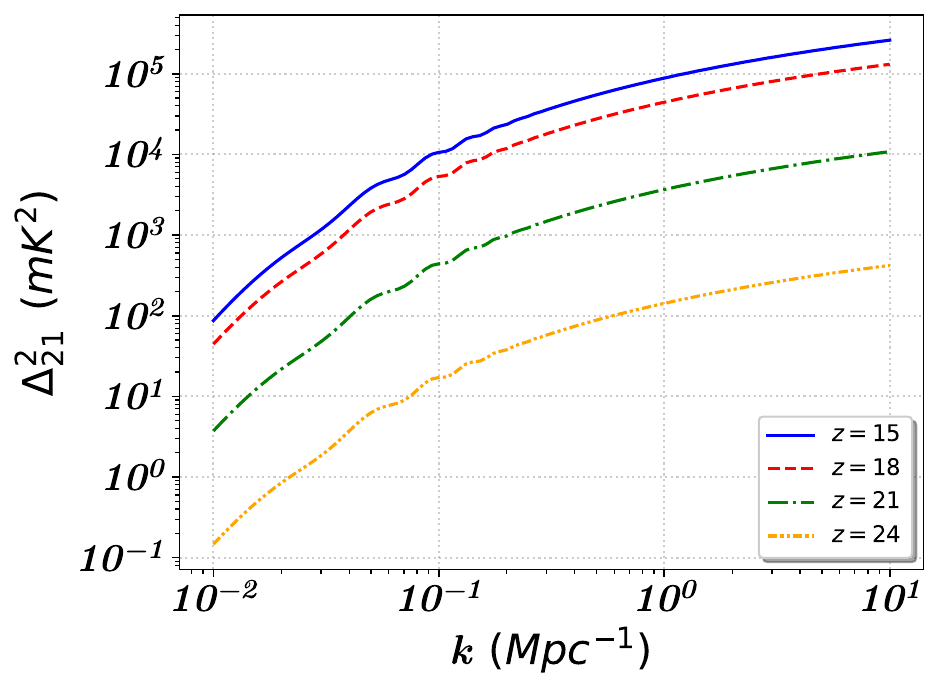}
    \caption{ Momentum profiles wrt $z$ for $B_0 = 0.0094$ nG}
    \label{fig:P21_B0fixed_z_varry}   
    \end{subfigure}
    \begin{subfigure}{.49\textwidth}
    \includegraphics[width=\textwidth]{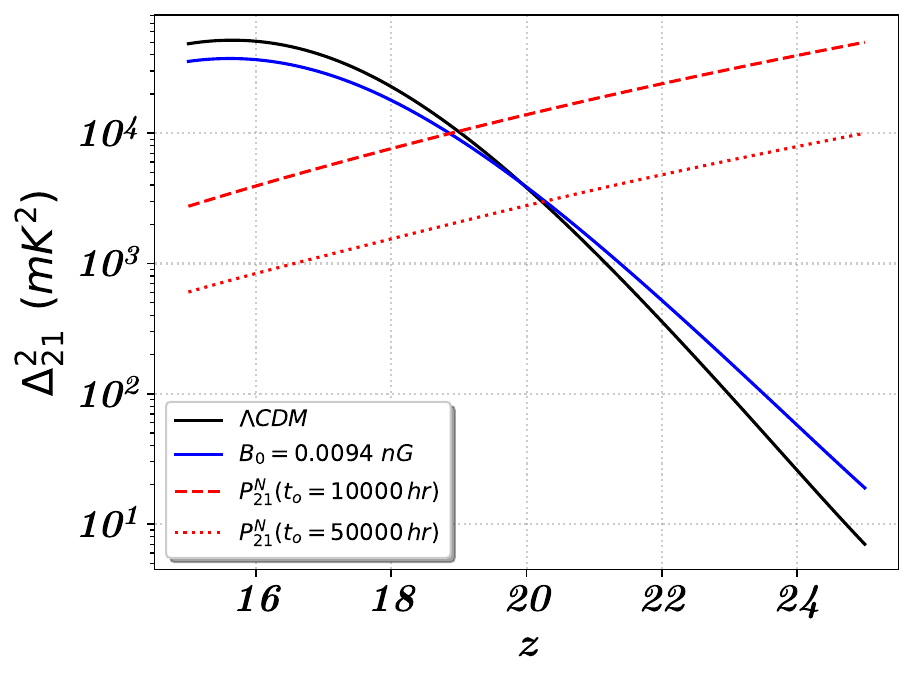}
    \caption{ Redshift profiles wrt $B_0$ at $k = 0.7$ Mpc$^{-1}$}
    \label{fig:P21_B0fixed_k_fixed}   
    \end{subfigure}
    \begin{subfigure}{.49\textwidth}
    \includegraphics[width=\textwidth]{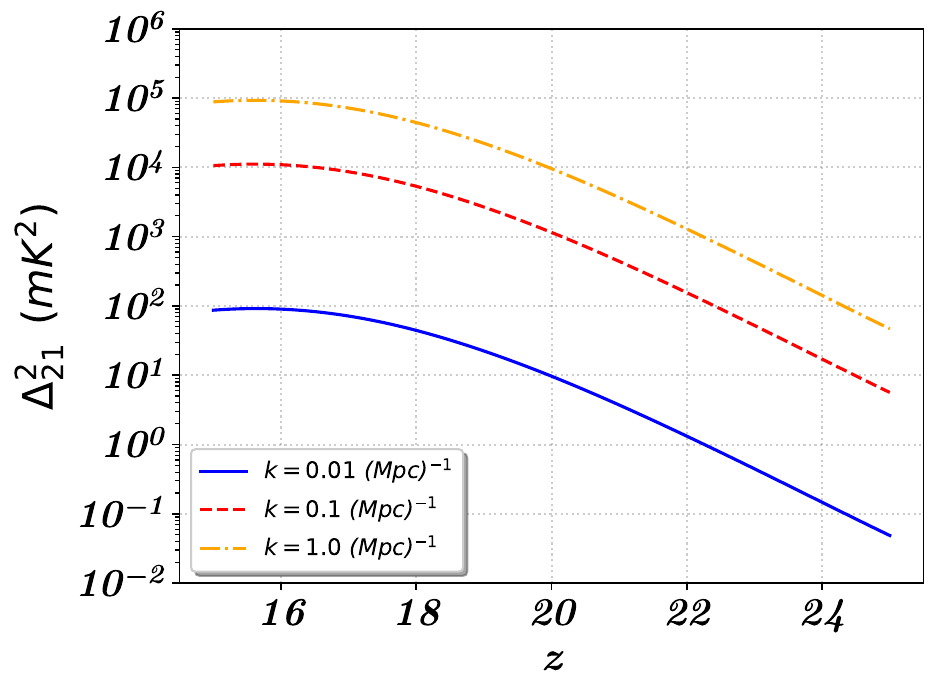}
    \caption{ Redshift profiles wrt $k$ for $B_0 = 0.0094$ nG}
    \label{fig:P21_B0fixed_k_varry}   
    \end{subfigure}
    \caption{ 21-cm power spectra shown with the total SKA-Low noise power spectrum (considering cosmic variance, thermal noise, and instrumental noise). In the upper row, the variation of the 21-cm power spectrum w.r.t. wave-number $k$ has been depicted, whereas the lower panel represents its variation w.r.t. redshift ($z$). In the left column, the noise spectra for observation times $10000$ hours and $50000$ hours are displayed, and the power spectra have been shown \textit{vis-\`{a}-vis} the $\Lambda$CDM scenario. We assume $\nb=-2.95$ and $\xi=20\%$ throughout.}
    \label{fig:P_21}
\end{figure*}

The $21$-cm power spectrum exhibits a strong dependence on the global signal, indicating that $P_{21}(k;z)$ is enhanced where $\deltat(z)$ shows its minimum. Since a non-helical PMF raises the IGM temperature, the dip in the global signal becomes shallower. Hence, the amplitude of the 21-cm power spectrum decreases upon increasing the magnetic field strength (for some fixed values of $\nb$ and $\xi$), which can be clearly seen in Fig.~\ref{fig:P21_noise_zfixed}. On the other hand, for a given set of parameter values, the depth of the global signal is maximised near $z\sim 16$, which results in an enhancement of power in this redshift region as illustrated in Fig.~\ref{fig:P21_B0fixed_z_varry}.

\section{Detection prospects at SKA-Low}
\label{sec:noise}
The SKA-Low is a phased array of simple dipole antennae covering frequencies ranging from $350$ MHz to $50$ MHz (\textit{i.e.} $z=4$ to $z=28$)~\footnote{\url{https://www.skao.int/en/explore/telescopes/ska-low}}. Since we are interested in the frequency range associated with EDGES ($z\sim\, 16-19$), the SKA-Low projection is thus essential for probing the relevant parameters.

\subsection{Noise modelling for SKA-Low}
\label{subsec:noise_ska}
We first elucidate the various sources of noise associated with the observation of the 21-cm power spectrum at SKA-Low. Our investigation takes into account three crucial sources of error that may contaminate the signal: the inherent statistical uncertainty known as \textit{cosmic variance}, \textit{thermal noise}, and the \textit{instrumental noise} of the experimental configuration.

\subsection*{Cosmic variance}
Cosmic variance (CV) is an intrinsic source of error due to the availability of only one realisation of the Universe. For 21-cm power spectrum observations, the CV noise is measured in terms of the bin-centred wavenumber $k$ and frequency $\nu$, with bandwidths $\Delta(\ln k)$ and $\Delta \nu$. In our case, we may take $\Delta(\ln k)=1$ and $\Delta \nu = 5$ MHz. Considering these, the CV noise can be expressed as~\cite{Mondal:2015oga}
\begin{eqnarray}\label{eq:CV_noise}
    P^{\rm N}_{\rm CV}(k;\nu) = \frac{2\pi P_{21}(k,\nu)}{\sqrt{V(\nu)k^3\Delta(\ln k)}},
\end{eqnarray}
where, to calculate $P_{21}(k;\nu)$, we have taken a $\mu$-averaged value of $P_{21}(\bold{k};z)$ (Eq.~\eqref{eq:21cmpowspecexpr}) and transformed it to frequency space via the relation $\nu = 1420/(1+z)$ MHz. In the expression above, $V(\nu)$ indicates the survey volume for the bin-centered frequency $\nu$ with bandwidth $\Delta\nu$, and be expressed as $V(\nu)=\Omega_{\!_{\rm FOV}}r_{\nu}^2\Delta r_{\Delta \nu}$. Here, $r_{\nu}$ represents the comoving distance for the bin-centered frequency $\nu$, $\Delta r_{\Delta \nu}$ is the comoving length interval corresponding to the bandwidth $\Delta\nu$, and $\Omega_{\!_{\rm FOV}} = \left(0.21(1+z){\rm m}\right)^2/A_{\rm eff}$ is the field of view (FOV) with an effective collecting area $A_{\rm eff} = (10^6/450)\, {\rm m^2}$~\cite{2013ExA....36..235M}.

\subsection*{Thermal noise}
Thermal noise is also measured in terms of the bin-centered variables $k$ and $\nu$, and can be expressed as~\cite{2013ExA....36..235M}
\be\label{eq:thermal_noise}
    P^{\rm N}_{\rm therm}(k;\nu) = \sqrt{\frac{4\, k^3\,V(\nu)}{\pi^2\,\Delta(\ln k)}}\, \frac{T_{\rm sys}^2}{\Delta\nu\, t_{\rm ini}}\, \frac{A_{\rm core}}{N_{\rm st}A_{\rm eff}},
\ee
where, for SKA-Low, the core area is given by $A_{\rm core} = 500^2$ m$^2$ and the total number of stations is $N_{\rm st} = 450$~\cite{2013ExA....36..235M}. The system temperature $T_{\rm sys}$ can be expressed as~\cite{deOliveira-Costa:2008cxd}
\begin{eqnarray}\label{eq:tsys}
    T_{\rm sys} = 180 \times \left(\frac{\nu}{180\,\text{MHz}}\right)^{-2.6}\, \text{K}.
\end{eqnarray}
For the bandwidths $\Delta(\ln k)$ and $\Delta\nu$, we have used the same values as for the CV noise.

\subsection*{Instrumental noise}
The instrumental noise for SKA-Low can be expressed in terms of the sky-covering fraction ($f_{\rm cover}$) and the observational period ($t_o$) as~\cite{Zaldarriaga:2003du,Tegmark:2008au,Modak:2021zgb}
\begin{eqnarray}\label{eq:ins_noise}
    P^{\rm N}_{\rm ins} (z) = \frac{\pi\, T_{\rm sys}^2}{f_{\rm cover}^2 \,t_o} d_{A}^2(z) y_{\nu}(z) \frac{\lambda^2(z)}{D_{\rm base}^2}\:,
\end{eqnarray}
where $\lambda(z)$ represents the $21$-cm transition wavelength at redshift $z$, and the angular diameter distance can be expressed as $d_{A}(z)=r_{z}/(1+z)$, with $r_{z}$ being the comoving distance at redshift $z$. On the other hand, $y_{\nu}$ is the conversion function from frequency ($\nu$) to $k_{\parallel}$, which can be expressed as~\cite{Tegmark:2008au}
\begin{eqnarray}
    y_{\nu} = 18.5\times\sqrt{\frac{1+z}{10}}\,\,\text{Mpc/MHz}.
\end{eqnarray}

For SKA-Low, we have considered a baseline of $D_{\rm base}=1$ km and sky-covering fraction $f_{\rm cover}\approx 0.0091$. Furthermore, we carry out our SNR analysis for two benchmark observation times $t_o = 10000$ hours and $t_o = 50000$ hours. The total resulting noise power spectrum can then be written as
\be\label{eq:total_noise}
   P^{\rm N}_{21}(k;z) = P^{\rm N}_{\rm CV}(k;z) + P^{\rm N}_{\rm therm}(k;z) + P^{\rm N}_{\rm ins}(z),
\ee
where the CV and thermal noises are expressed in redshift space by using $\nu = 1420/(1+z)$ MHz. From Fig.~\ref{fig:P_21}, we observe that CV is the dominant source of noise on large scales, while instrumental noise becomes important at smaller scales. Moreover, the instrumental noise is strong enough to dominate over thermal noise across all scales. Fig.~\ref{fig:P_21} additionally depicts that the instrumental noise scales inversely with the observational time, thereby raising the SNR at smaller scales for longer mission durations.

\subsection{Estimation of signal-to-noise ratio (SNR)}
\label{subsec:snr}
\begin{figure*}[!ht]
    \centering
    \begin{subfigure}{.49\textwidth}
    \includegraphics[width=\textwidth]{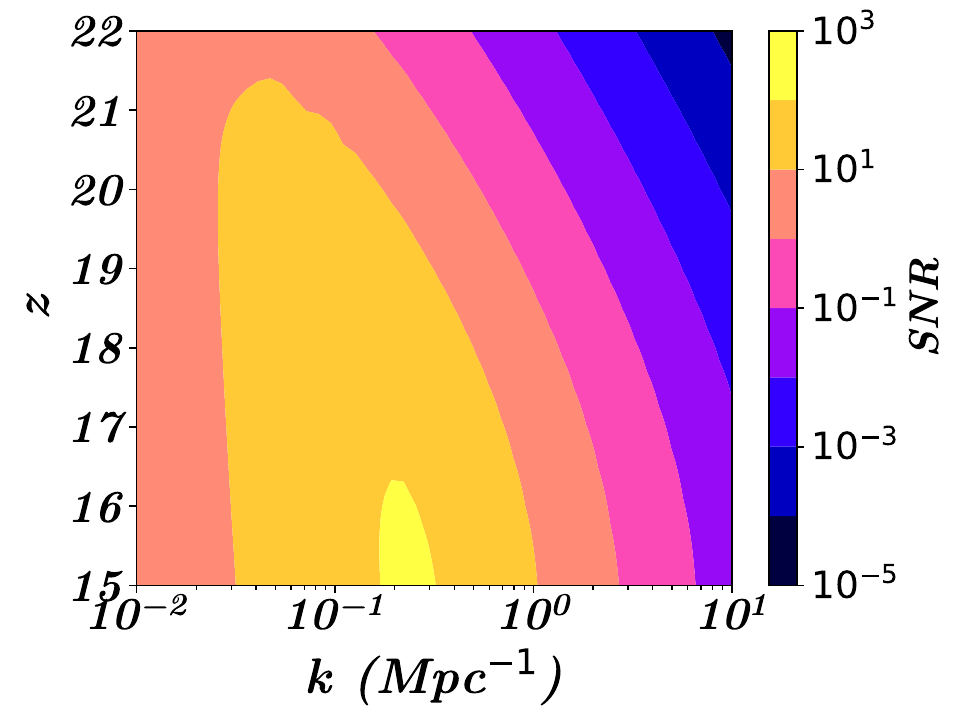}
    \caption{ $t_o = 10000$ hours}
    \label{fig:snr_10000}   
    \end{subfigure}
    \begin{subfigure}{.49\textwidth}
    \includegraphics[width=\textwidth]{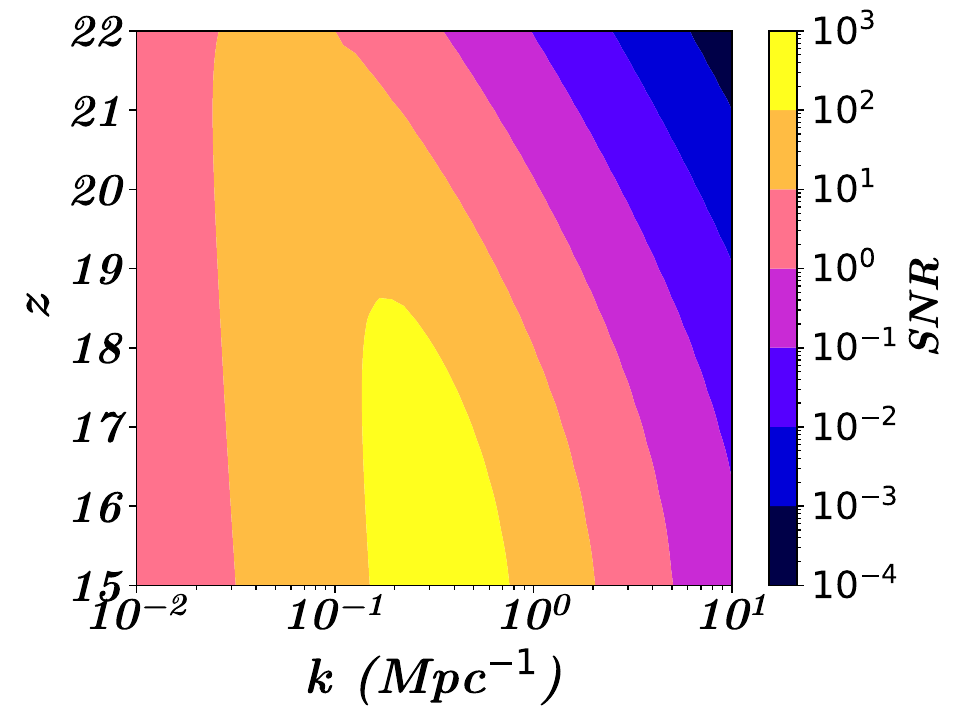}
    \caption{ $t_o = 50000$ hours}
    \label{fig:snr_50000}
    \end{subfigure}
    \caption{ Dependence of the 21-cm power spectrum SNR at SKA-Low on the wavenumber ($k$) and the redshift ($z$), considering $B_0 = 0.00942$ nG and $\nb = -2.95$. The left plot is for $t_o = 10000$ hours, whereas the right plot is for $t_o = 50000$ hours. The noise model incorporates cosmic variance, thermal noise, and the instrumental noise.}
    \label{fig:snr}
\end{figure*}

For our purpose, the SNR may be defined as an explicit function of both $k$ and $z$ as
\be\label{eq:snr}
   {\rm SNR} \equiv \frac{P_{21}(k;z)}{P_{21}^{\rm N} (k;z)}\:,
\ee
to arrive at which, we have first performed $\mu$-averaging of $P_{21}(\bold{k};z)$ to obtain $P_{21}(k;z)$ (\textit{viz.} Eq.~\eqref{eq:21cmpowspecexpr}). For a given set of $\{B_0,\nb,\xi\}$, the SNR is found to depend strongly on both $k$ and $z$. This is shown in Fig.~\ref{fig:snr}, which represents the SNR of the 21-cm power spectrum at SKA-Low in the $k$-$z$ plane corresponding to the benchmark parameter values $B_0=0.00942$ nG, $\nb=-2.95$, and $\xi=15\%$. As visible in the plot, the SNR peaks close to $z\sim 16$ and in the range $k\in(0.1,1.0)$. The origin of this feature can be traced back to Fig.~\ref{fig:T_21_2.95}, which displays that the global signal has a trough around $z\sim16$, which causes an enhancement in the power spectrum near $z\sim 16$ (\textit{viz.} Fig.~\ref{fig:P21_B0fixed_z_varry}) and results in a corresponding rise in the SNR. As for the $k$-range, the noise level is significant both at small $k$ (due to CV) and at large $k$ (due to instrumental noise) as might be gleaned from \textit{viz.} Fig.~\ref{fig:P21_noise_zfixed}, which results in an enhancement of the SNR in the mid-$k$ region. Additionally, it should be noted that large-scale foreground contamination may  hinder the detection of the power spectrum at very small values of $k$ to some extent, which fall towards the extreme left edges of the plots in Fig. \ref{fig:snr}. We also compare the SNR for $t_o=10000$ hours (Fig.~\ref{fig:snr_10000}) and $t_o=50000$ hours (Fig.~\ref{fig:snr_50000}), which shows that the SNR increases nearly one order of magnitude for the longer observational period. This manifests especially at large $k$ due to lesser instrumental noise levels, as evident from Eq.~\eqref{eq:ins_noise}.

\begin{figure*}[!ht]
    \centering
    \begin{subfigure}{.49\textwidth}
    \includegraphics[width=\textwidth]{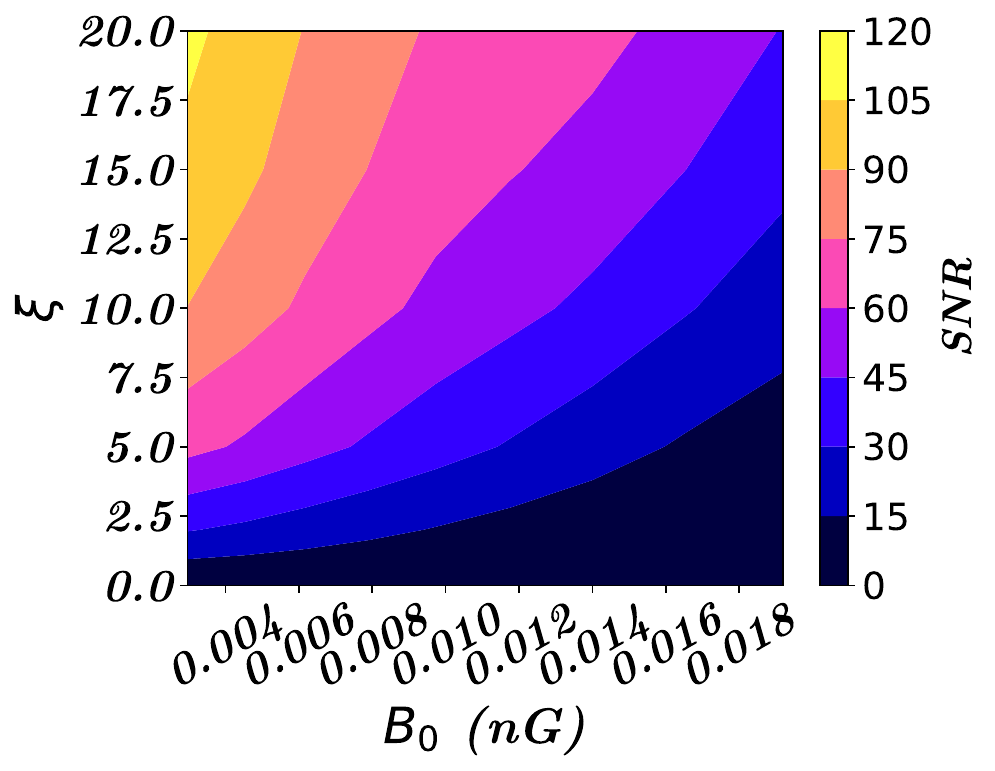}
    \caption{ $\nb = -2.75$}
    \label{fig:snr_10000_275}   
    \end{subfigure}
    \begin{subfigure}{.49\textwidth}
    \includegraphics[width=\textwidth]{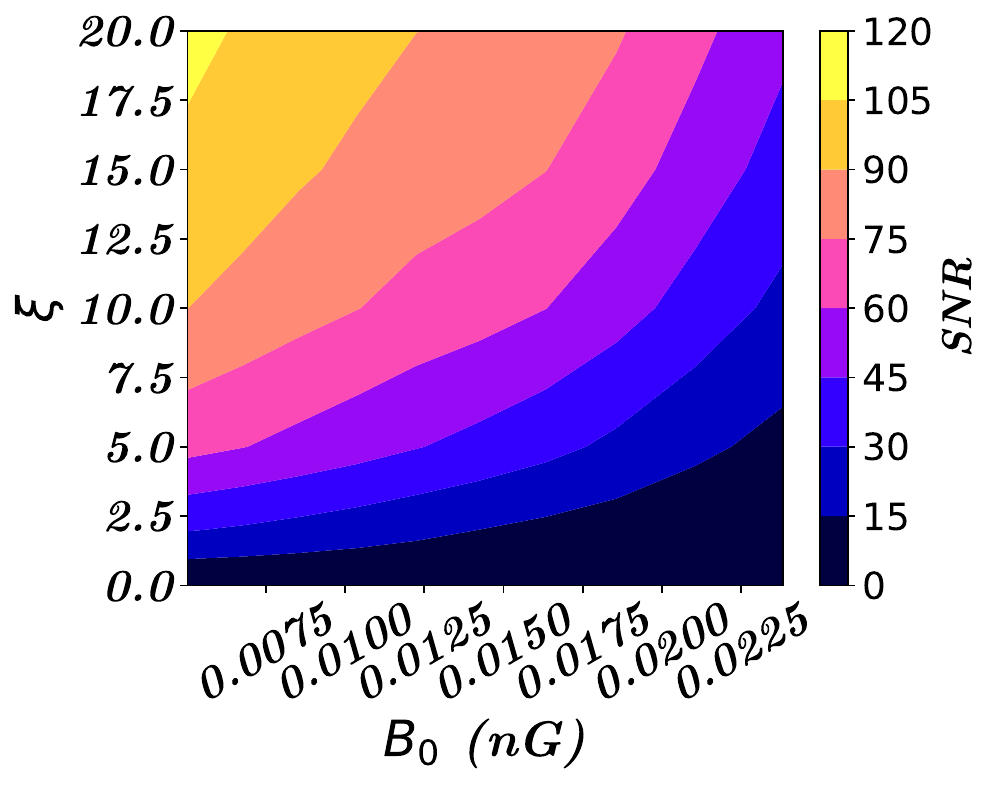}
    \caption{ $\nb = -2.95$}
    \label{fig:snr_10000_295}
    \end{subfigure}
    \caption{ Dependence of the 21-cm power spectrum SNR at SKA-Low on the present day magnetic field strength ($B_0$) and the excess radio background parameter ($\xi$), calculated at $k = 0.3$ Mpc$^{-1}$ and $z = 16$. The left plot is for $\nb = -2.75$, whereas the right plot is for $\nb = -2.95$. The values of $\xi$ represented in the figure are in percentage fraction. The noise model incorporates cosmic variance, thermal noise, and the instrumental noise.}
    \label{fig:snr_B0_xi}
\end{figure*}

The dependence of SNR on the parameters $\{B_0,\nb,\xi\}$ themselves, on the other hand, is depicted in Fig.~\ref{fig:snr_B0_xi}, which displays the SNR in the $B_0$-$\xi$ plane for the two values of $\nb=-2.75$ (Fig.~\ref{fig:snr_10000_275}) and $\nb=-2.95$ (Fig.~\ref{fig:snr_10000_295}), corresponding to fixed $k=0.3$ Mpc$^{-1}$ and $z=16$. The SNR is found to be maximised for the least value of $B_0$ and the highest value of $\xi$. This is in tune with the fact that the PMF heats the gas and attenuates the global signal for larger $B_0$, which, in turn, dampens the 21-cm power spectrum. On the other hand, a stronger excess radio background increases the depth of the global signal, resulting in an increment of power.

\subsection{Fisher forecast analysis}
\label{subsec:fisher_results}
While the SNR provides a decent idea of the parameter range needed for robust detection, the standard Fisher forecast methodology (with the noise model described in Sec.~\ref{subsec:noise_ska}) is implemented in this section to assess the experimental capability of SKA-Low in constraining the PMF parameters ($B_0$ and $n_{\!_{B}}$) jointly with the excess radio background parameter ($\xi$). For maximum likelihood analysis, the standard prescription is to maximise the log-likelihood function ($\mathcal{L}$), in terms of which the Fisher matrix is defined as~\cite{Tegmark:1996bz}
\be\label{eq:fisher_basic}
   F_{ab} = \left\langle-\,\frac{\partial^2 \mathcal{L} (\bold{\Theta})}{\partial \Theta_a \partial \Theta_b}\right\rangle,
\ee
where $\{\Theta_i\}$ indicates the vector of model parameters, with angular brackets representing the angular average over the observational data. 

For SKA-Low which is expected to probe $3\lesssim z\lesssim28$, we consider a mock cosine-averaged $z$-binned 1D power spectrum dataset at a constant $k$-slice. Specifically, the $\mu$-averaged $P_{21}(k,z)$ at a fixed benchmark value of $k=0.3$ Mpc$^{-1}$ has been considered as a function of the redshift, with $z$-binnings in accordance with the instrumental specifications of SKA-Low. This is similar to a tomographic study of the 21-cm power spectrum with SKA-Low \cite{Furlanetto:2004zw,Mao:2008ug,Chen:2016zuu,Gillet:2018fgb} at a fixed scale. In terms of the noise model, the log-likelihood function for SKA-Low then involves a sum over all the redshift bins, and can be expressed as~\cite{Munoz:2016owz}
\be\label{eq:log-likelihood}
   \mathcal{L}(\bold{\Theta}; k,z) = \dfrac { f_{\rm sky}} 2 \sum\limits_{i} \int_{-1}^{1} \!\!\!\mathrm d \mu \: \frac{\left(P_{21}(k,z_i,\mu,\bold{\Theta}) - P_{21}^{\rm fid}(k,z_i,\mu,\bold{\Theta}_{\rm fid}) \right)^2}{\left( P_{21}(k,z_i,\mu,\bold{\Theta}) + P_{21}^N(z_i)\right)^2}.
\ee
Using this in Eq.~\eqref{eq:fisher_basic}, the Fisher matrix is given by
\be\label{eq:fisher}
   F_{a b}(z) = \dfrac { f_{\rm sky}} 2 \sum\limits_{i} \int_{-1}^{1} \!\!\!\mathrm d \mu \left [ P_{21}(\bold{k},z,\bold{\Theta}) + P_{21}^N(z)\right]^{-2}\dfrac{\partial P_{21}(\bold{k},z, \bold{\Theta})}{\partial \Theta_a}\dfrac{\partial P_{21}(\bold{k},z,\bold{\Theta})}{\partial \Theta_b}.
\ee
For the present analysis, a sky fraction of $f_{sky}=0.72$ has been assumed (corresponding to a cosmological mask~\cite{Villaescusa-Navarro:2016kbz}), alongside a runtime of $t_o=50000$ hours. Equipped with Eq.~\eqref{eq:fisher}, we now proceed to estimate the $1\sigma$ uncertainties on $\{B_0,\nb,\xi\}$ with SKA-Low.

\begin{figure*}[!ht]
    \centering
    \begin{subfigure}{0.49\textwidth}
    \includegraphics[width=\textwidth]{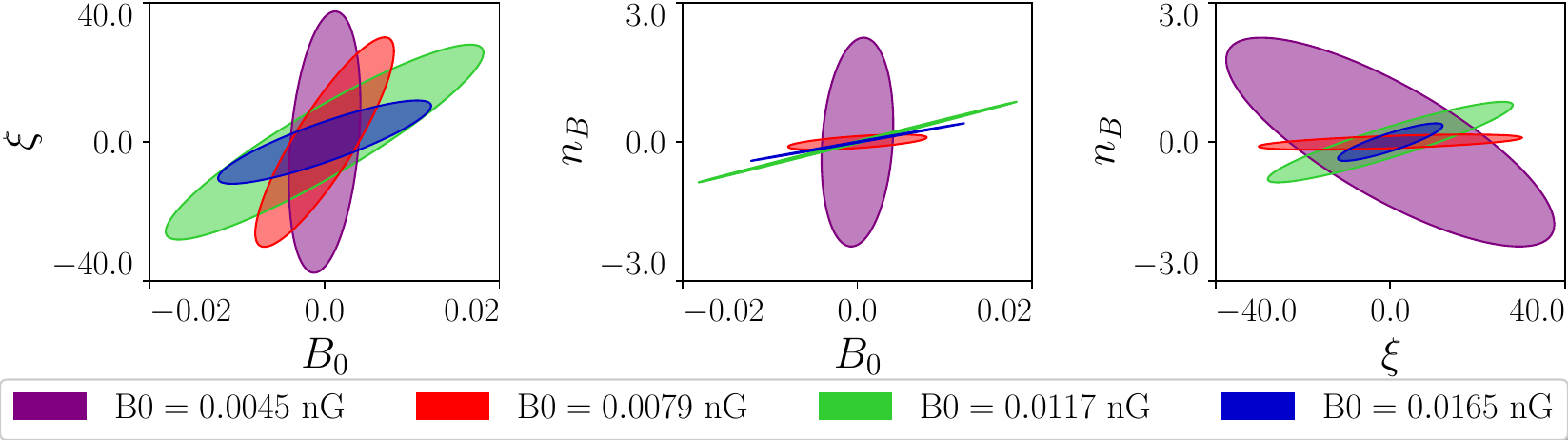}
    \caption{ $n_{\!_{B}} = -2.75$}
    \label{fig:sfisher_275}
    \end{subfigure}
    \begin{subfigure}{0.49\textwidth}
    \includegraphics[width=\textwidth]{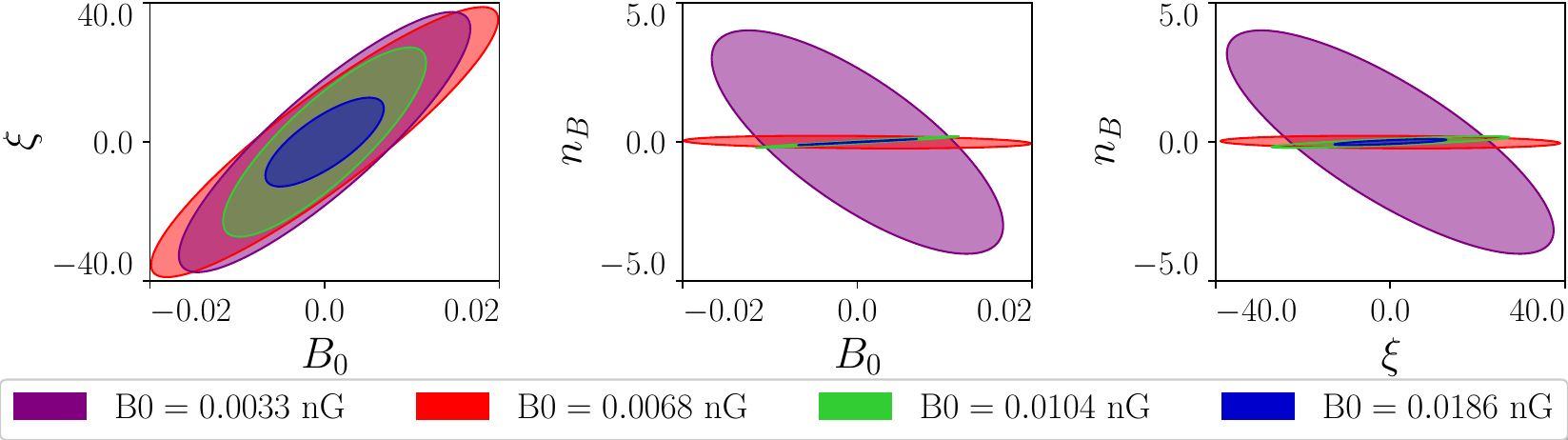}
    \caption{ $n_{\!_{B}} = -2.95$}
    \label{fig:sfisher_295}
    \end{subfigure}
    \caption{ Marginalised $1\sigma$ contours of the two-dimensional posterior distribution obtained from Fisher matrix analysis for SKA-Low. For this set of plots, we have considered $\xi=15\%$. To highlight the correlations, actual fiducial positions have not been displayed. The values of $\xi$ are in percentage fraction. The left set of plots is for $\nb=-2.75$, whereas the right one is for $\nb=-2.95$.}
    \label{fig:fisher_A}
\end{figure*}

\begin{figure*}[!ht]
    \centering
    \begin{subfigure}{0.49\textwidth}
    \includegraphics[width=\textwidth, trim={0 1.16cm 0 0},clip]{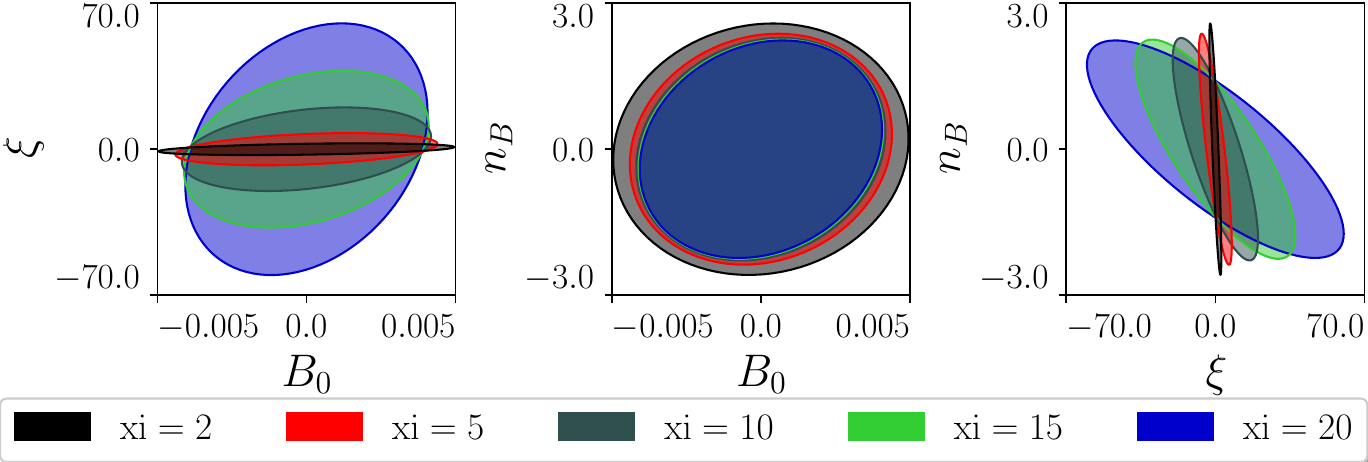}
    \caption{ For $B_0 = 0.0045$ nG}
    \label{fig:sfisher_30000_275}   
    \end{subfigure}\hfill
    \begin{subfigure}{0.49\textwidth}
    \includegraphics[width=\textwidth, trim={0 1.16cm 0 0},clip]{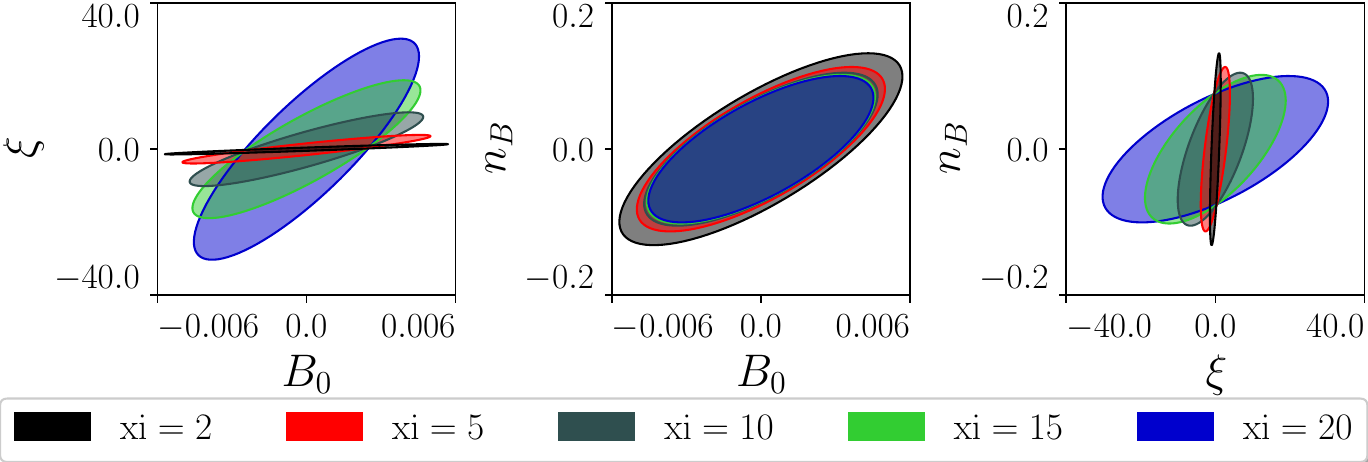}
    \caption{ For $B_0 = 0.0079$ nG}
    \label{fig:sfisher_50000_275}  
    \end{subfigure}\hfill
    \begin{subfigure}{0.49\textwidth}
    \includegraphics[width=\textwidth, trim={0 1.16cm 0 0},clip]{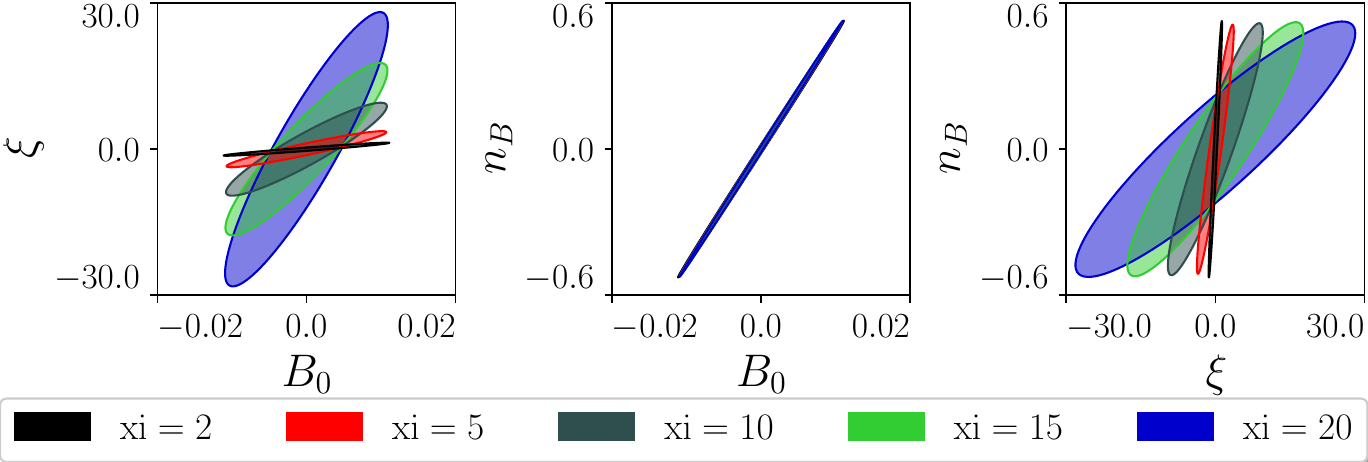}
    \caption{ For $B_0 = 0.0117$ nG}
    \label{fig:sfisher_70000_275}  
    \end{subfigure}
    \begin{subfigure}{0.49\textwidth}
    \includegraphics[width=\textwidth, trim={0 1.16cm 0 0},clip]{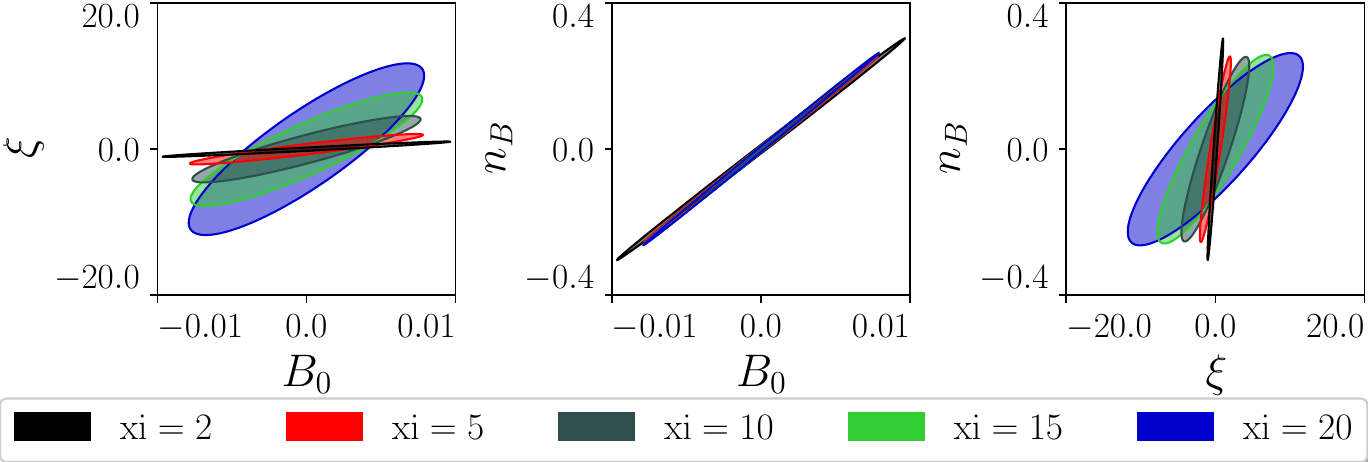}
    \caption{ For $B_0 = 0.0165$ nG}
    \label{fig:sfisher_90000_275}  
    \end{subfigure}
    \begin{subfigure}{0.49\textwidth}
    \includegraphics[width=9cm, trim={0 0 0 6.7cm},clip]{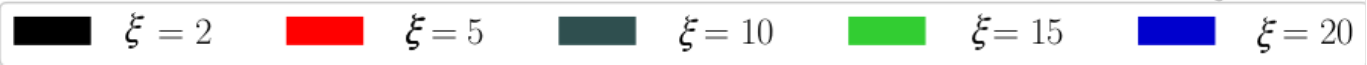}
    \end{subfigure}
    \caption{ Marginalised $1\sigma$ contours of the two-dimensional posterior distribution obtained from Fisher matrix analysis for SKA-Low, corresponding to $\nb=-2.75$. The values of $\xi$ quoted in the figure are in percentage fraction. To highlight the correlations, actual fiducial positions have not been displayed.}
    \label{fig:fisher_B_2.75}
\end{figure*}

\begin{figure*}[!ht]
    \centering
    \begin{subfigure}{.49\textwidth}
    \includegraphics[width=\textwidth, trim={0 1.16cm 0 0},clip]{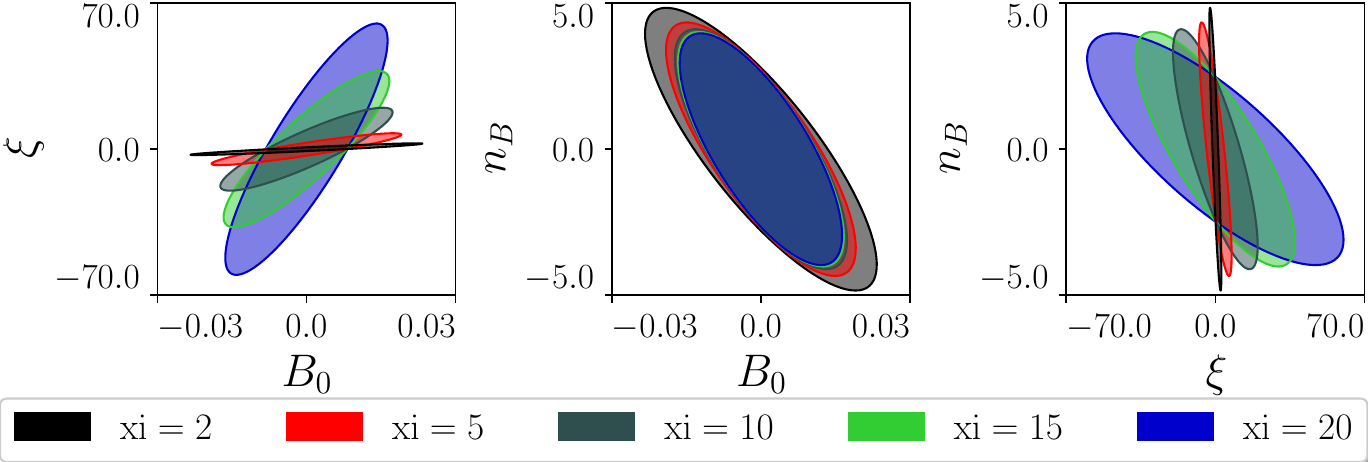}
    \caption{ For $B_0 = 0.0033$ nG}
    \label{fig:sfisher_10000_295}   
    \end{subfigure}\hfill
    \begin{subfigure}{.49\textwidth}
    \includegraphics[width=\textwidth, trim={0 1.16cm 0 0},clip]{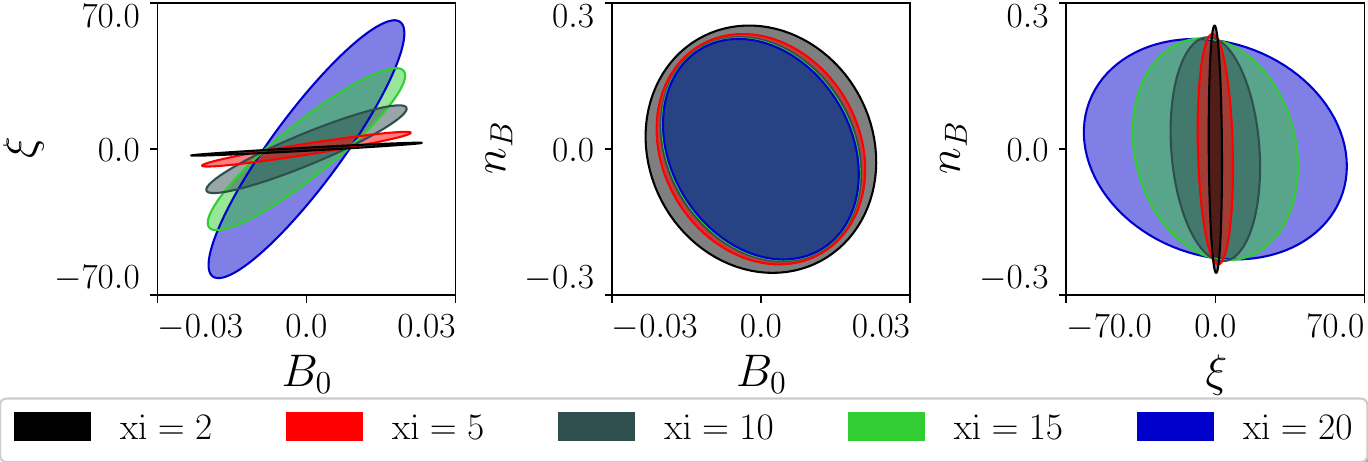}
    \caption{ For $B_0 = 0.0068$ nG}
    \label{fig:sfisher_20000_295}  
    \end{subfigure}
    \begin{subfigure}{.49\textwidth}
    \includegraphics[width=\textwidth, trim={0 1.16cm 0 0},clip]{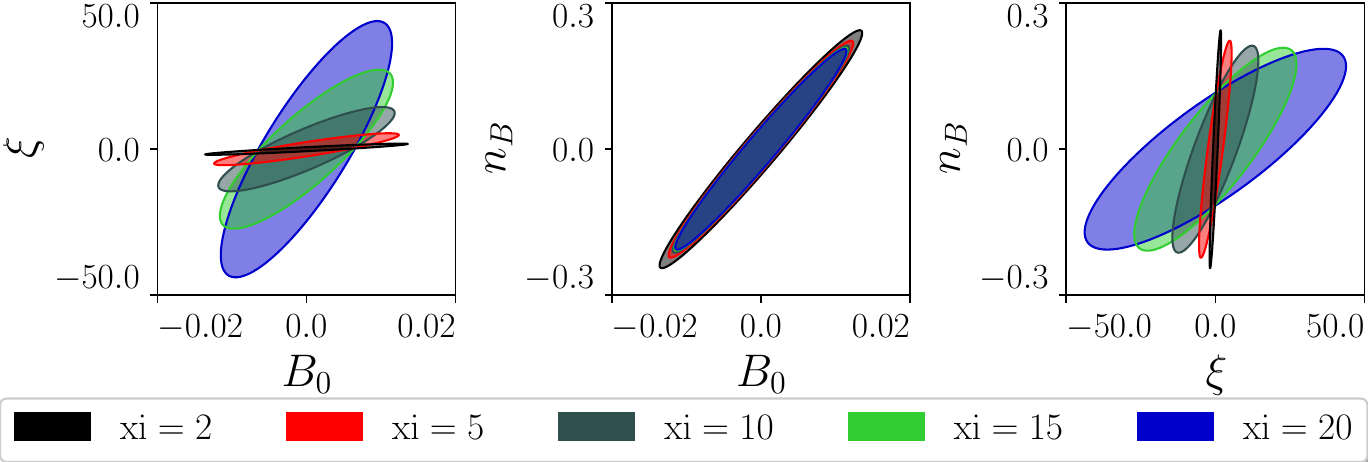}
    \caption{ For $B_0 = 0.0104$ nG}
    \label{fig:sfisher_30000_295}  
    \end{subfigure}\hfill
    \begin{subfigure}{.49\textwidth}
    \includegraphics[width=\textwidth, trim={0 1.16cm 0 0},clip]{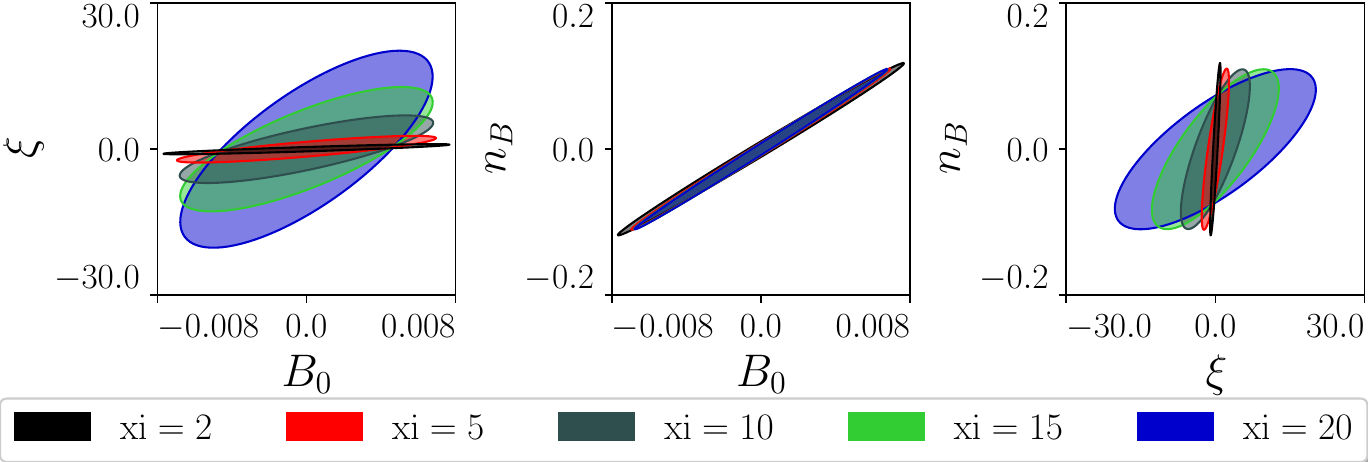}
    \caption{ For $B_0 = 0.0186$ nG}
    \label{fig:sfisher_50000_295}  
    \end{subfigure}
    \begin{subfigure}{.49\textwidth}
    \includegraphics[width=9cm, trim={0 0 0 6.7cm},clip]{Figures/xi_label.pdf}
    \end{subfigure}
    \caption{ Marginalised $1\sigma$ contours of the two-dimensional posterior distribution obtained from Fisher matrix analysis for SKA-Low, corresponding to $\nb=-2.95$. The values of $\xi$ quoted in the figure are in percentage fraction. To highlight the correlations, actual fiducial positions have not been displayed.}
    \label{fig:fisher_B_2.95}
\end{figure*}

Figs.~\ref{fig:fisher_A}, \ref{fig:fisher_B_2.75}, and \ref{fig:fisher_B_2.95} depict the predicted 2D posterior distributions (at $1\sigma$) for different fiducial values of the parameters (actual fiducial positions have not been displayed to compare the correlations). To plot these results, we have used the plotting library of the publicly available Fisher analysis code \texttt{CosmicFish}~\cite{Raveri:2016xof,Raveri:2016leq}. Fig.~\ref{fig:fisher_A} illustrates the comparisons among the uncertainties associated with different magnetic field strengths for given fiducial values of $\xi = 15\%$ and $\nb = -2.75$ ($\nb = -2.95$) in Fig.~\ref{fig:sfisher_275} (Fig.~\ref{fig:sfisher_295}). In the $B_0$-$\xi$ plane, the uncertainties turn out to be strongly positive correlated for $B_0=0.0117$ nG for $\nb=-2.75$, whereas the correlation decreases for both $B_0>0.0117$ nG and $B_0<0.0117$ nG. This feature is more prominent as the magnetic field becomes relatively more scale-dependent \textit{i.e.} for more positive values of $\nb$. On top of this, the uncertainty associated with $B_0=0.0117$ nG is larger. Furthermore, in the $B_0$-$\nb$ plane, $B_0=0.0117$ nG shows a strong positive correlation with maximum uncertainty among all the fiducial values considered. For $\nb=-2.95$, on the other hand, the dependence of the correlation strength in the $B_0$-$\xi$ plane on the fiducial value of $B_0$ is observed to be lesser, while similar flips in the correlation are observed at some critical value of $B_0$ in the $B_0-\nb$ and $\xi-\nb$ planes.

We have also compared the uncertainties associated with the different values of $\xi$ for several fiducial values of $B_0$ and $n_{\!_{B}}$, which have been depicted in Figs.~\ref{fig:fisher_B_2.75} (for $\nb=-2.75$) and \ref{fig:fisher_B_2.95} (for $\nb=-2.95$). Both of the figures represent that $B_0$ and $\xi$ are positively correlated for every $B_0$ and $\nb$, which is expected since they provide mutually antagonistic effects on $\deltat(z)$. Moreover, $B_0$ and $\nb$ are in strong positive correlation for relatively large magnetic field strengths, whereas the correlation vanishes for smaller fields and ultimately switches to negative for sufficiently small magnetic fields. The latter feature is more prominent as $n_{\!_{B}}\to-3$, which is why Fig.~\ref{fig:fisher_B_2.95} shows a transition from positive to negative correlation in the $B_0-\nb$ plane as $B_0$ decreases. The $\xi-\nb$ plane also exhibits the same transition as the magnetic field strength falls.

\begin{figure*}[!ht]
    \centering
    \begin{subfigure}{.32\textwidth}
    \includegraphics[width=\textwidth]{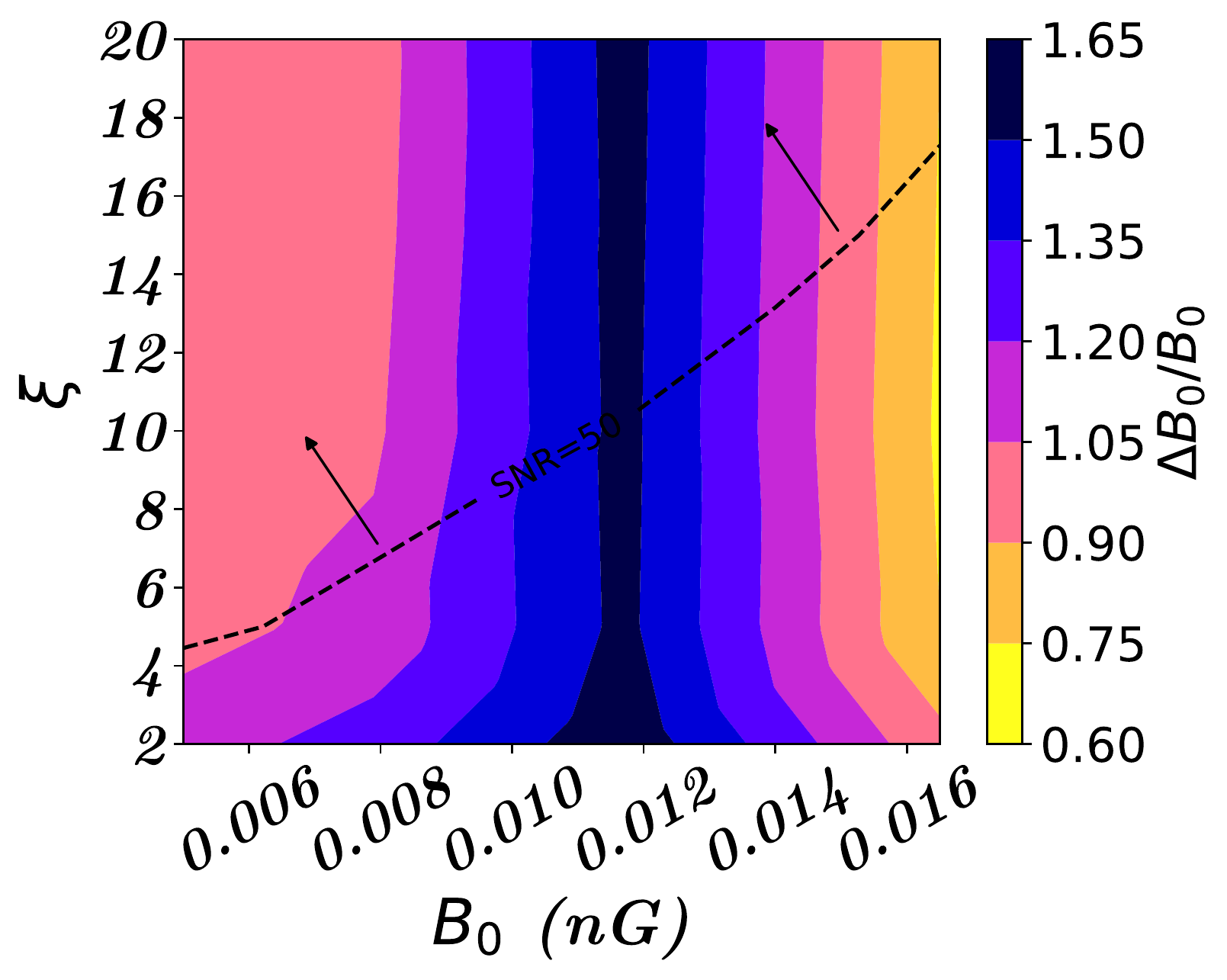}
    \caption{}
    \label{fig:fisher_275_b0_snr}   
    \end{subfigure}
    \begin{subfigure}{.32\textwidth}
    \includegraphics[width=\textwidth]{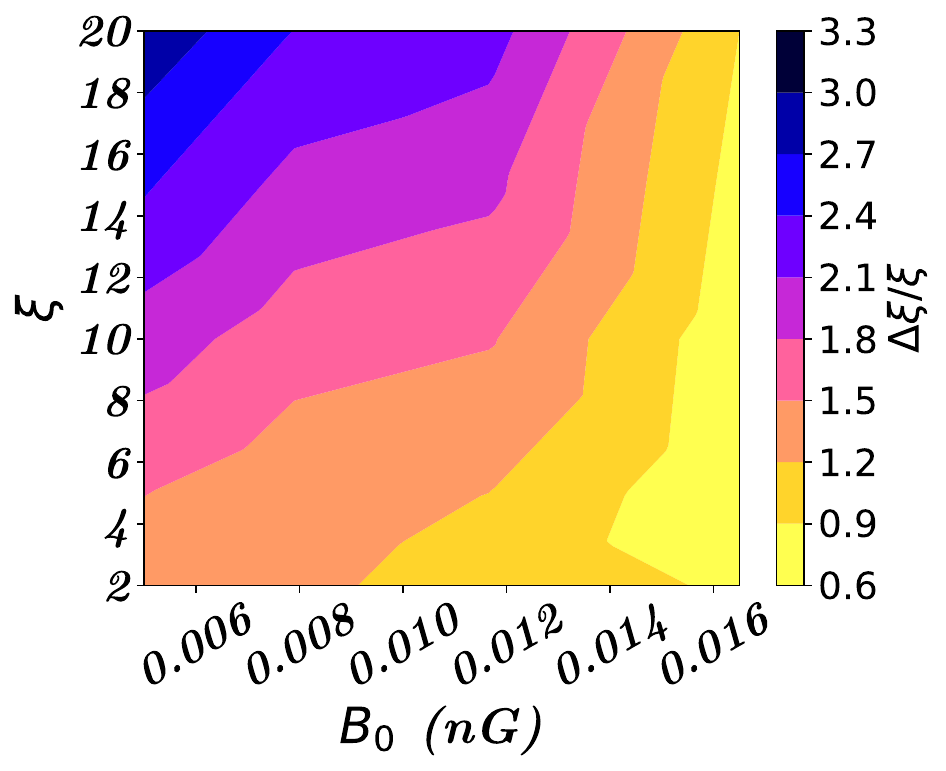}
    \caption{}
    \label{fig:fisher_275_xi_snr}   
    \end{subfigure}
    \begin{subfigure}{.32\textwidth}
    \includegraphics[width=\textwidth]{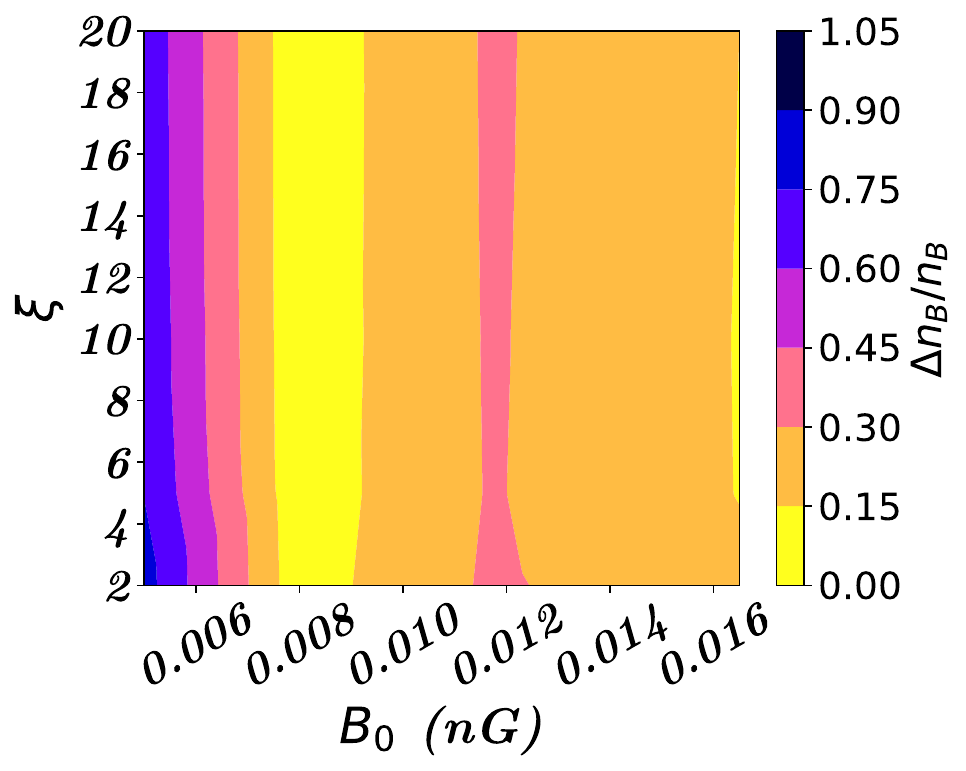}
    \caption{}
    \label{fig:fisher_275_nb_snr}   
    \end{subfigure}
    \caption{ Variation of the relative errors on the $B_0-\xi$ plane, corresponding to $\nb=-2.75$ and $t_o=50000$ hours. The values of $\xi$ quoted in the figure are in percentage fraction.}
    \label{fig:fisher_SNR_2.75}
\end{figure*}

\begin{figure*}[!ht]
    \centering
    \begin{subfigure}{.32\textwidth}
    \includegraphics[width=\textwidth]{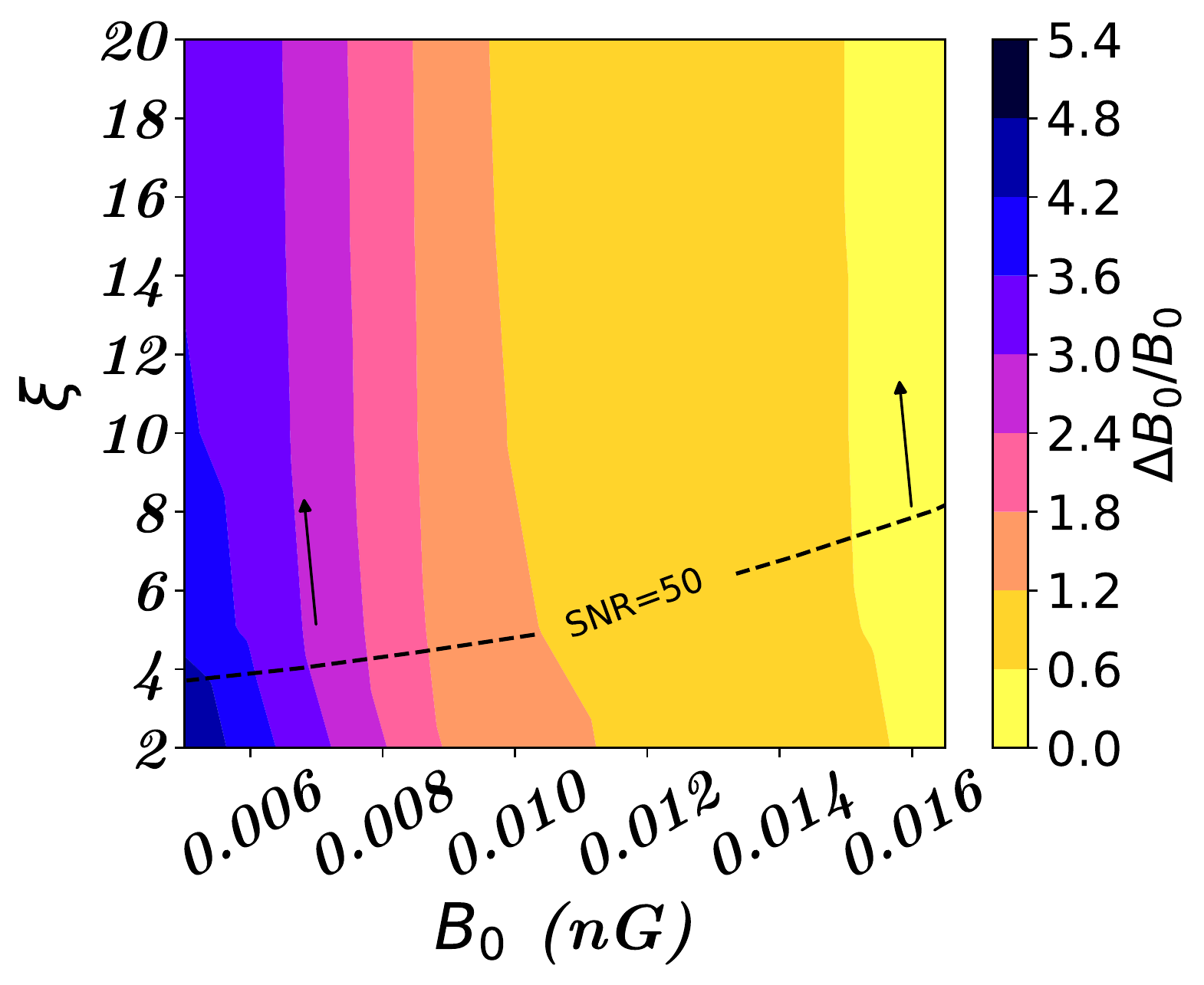}
    \caption{}
    \label{fig:fisher_295_b0_snr}   
    \end{subfigure}\hfill
    \begin{subfigure}{.32\textwidth}
    \includegraphics[width=\textwidth]{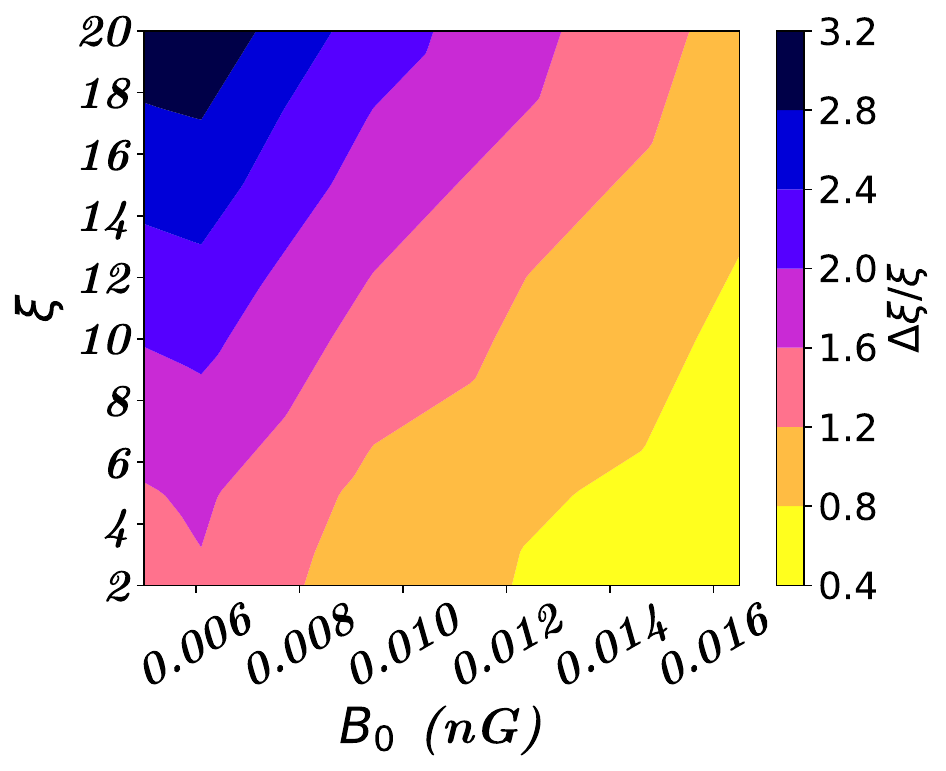}
    \caption{}
    \label{fig:fisher_295_xi_snr}   
    \end{subfigure}
    \begin{subfigure}{.32\textwidth}
    \includegraphics[width=\textwidth]{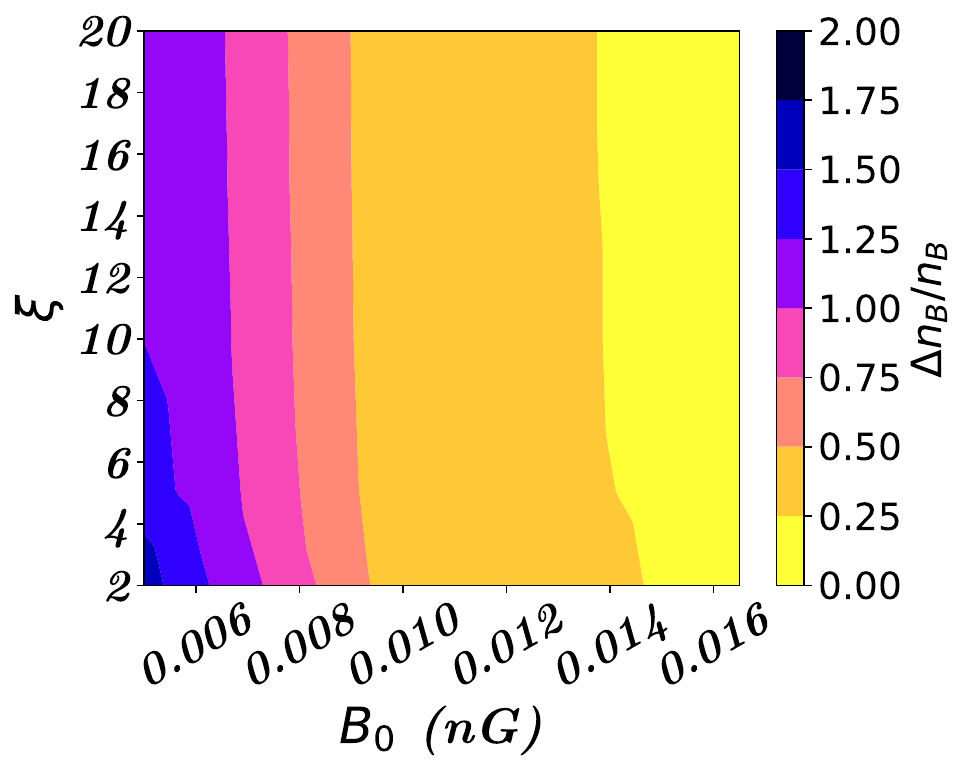}
    \caption{}
    \label{fig:fisher_295_nb_snr}   
    \end{subfigure}
    \caption{ Variation of the relative errors on the $B_0-\xi$ plane, corresponding to $\nb=-2.95$ and $t_o=50000$ hours. The values of $\xi$ quoted in the figure are in percentage fraction.}
    \label{fig:fisher_SNR_2.95}
\end{figure*}

Finally, as Fisher results are inherently dependent on the choice of fiducials, the behaviour of the relative $1\sigma$ uncertainties is displayed in Figs.~\ref{fig:fisher_SNR_2.75} (for $\nb=-2.75$) and \ref{fig:fisher_SNR_2.95} (for $\nb=-2.95$) as functions of the fiducial values. These contour maps indicate that particular regions of the parameter space, where the relative $1\sigma$ uncertainties associated with the various parameters are less than $10\%$, should fall within the detectable threshold of SKA-Low. 

\section{Summary and future directions}
\label{sec:conclusion}

In this article we have investigated the effects of PMF on the global 21-cm signal as well as on the 21-cm power spectrum in a thorough, consistent manner, and found out possible reflections on the relevant observations. Our work consists of two major parts. In the first part, we have investigated possible bounds on PMF parameters based on the global 21-cm signal observed by EDGES, by integrating both IGM heating effects (AD and DT) and PMF-induced modification of the matter power spectrum within a unified framework. The former controls only the vertical height of the global 21-cm trough, whereas the latter impacts the star formation history thereby reflecting on the redshift of the dip. These two effects are usually considered separately in the literature, resulting in their inherent limitations.  Our work presents a more comprehensive view by simultaneously addressing these two effects in light of the EDGES signal. The combined impact of both phenomena makes the behaviour of $\deltat(z)$ depart significantly from scenarios where either effect is considered alone. Additionally, we have considered an excess radio background (characterized by $\xi$) on top of the PMF parameter space ($B_0$ and $\nb$), with the former providing the essential IGM cooling effect. Our work reveals that the upper bound on $B_0$ for a nearly scale-invariant, non-helical PMF should be typically around $\mathcal{O}(10^{-2})$ nG for different choices of the parameters $\nb$ and $\xi$, which is tighter than the previously reported values. We also emphasize that due to AD and DT, the evolution of the magnetic field amplitude should deviate from the trivial adiabatic evolution given by $B(z)=B_0(1+z)^2$, which has been assumed in some of the earlier works. This issue has been properly addressed through our adopted methodology. Interestingly, as $B_0$ starts increasing from zero, a marginal leftward shift of the global 21-cm trough to lower redshifts is observed, whereafter it visibly moves rightward upon further increasing the field strength. The precise reason is that the SFRD curves for distinct values of $B_0$ may exhibit crossings with each other at redshifts higher than the interval relevant to EDGES. This, in turn, leads to interesting trends in the correlations among the various parameters.

In the second half of our work, we have turned towards the prospects of detection of effects of PMF on the the 21-cm power spectrum from the future SKA-Low facility. The magnitude of the $\deltat$ dip was found to have significantly influenced the amplitude of the 21-cm power spectrum. With the minimum of $\deltat(z)$ occurring at $z\sim 16-19$, the power at a constant $z$ value (chosen to be $z=16$ for demonstration) gets dampened upon increasing $B_0$ for a fixed value of $\xi$, as it leads to a shallower dip. On the other hand, the power at a fixed $z$ rises with increasing $k$ for any given value of $B_0$, $\nb$, and $\xi$. Considering possible sources of noise for SKA-Low, it was found that, away from both the high and low-$k$ ends, the $k\in(0.1,1.0)$ Mpc$^{-1}$ zone is mostly signal dominated, resulting in an SNR as high as $\mathcal{O}(10^2)$ at times. As a function of the fiducial parameter values and at some fixed scale and redshift, we observe the SNR to peak for smaller values of $B_0$ and larger values of $\xi$, as expected owing to their antagonistic effects on the depth of the 21-cm trough. Further, to estimate the projected uncertainties associated with the three parameters $\{B_0,\nb,\xi\}$ at SKA-Low, a joint Fisher forecast analysis has been carried out for all of these parameters. Our results shed light on the expected $1\sigma$ errors as well as on the mutual correlations among these three parameters, which show certain interesting trends and may be attributed to various physical processes involved. The error forecasts lead to encouraging results, wherefore it seems possible to constrain particular regions of the parameter space with a relative error $\lesssim10\%$.

Before concluding, we reiterate that our results correspond to a fixed set of benchmark values of the astrophysical parameters $f_*$, $f_X$, and $f_{\rm heat}$. In particular, the X-ray heating parameter remains poorly constrained as of now, and may vary several orders of magnitude between $10^{-4}$ and $10^3$ based on presently available datasets \cite{HERA:2021noe}. Larger values of the $f_{\rm heat}f_X$ product should lead to significantly tighter upper bounds on $B_0$ based on the lateral shift of the 21-cm trough to higher redshifts, as may be seen from \eqref{eq:epsX}. Our obtained values thus come with an inherent note of caution, with any stronger comment in this regard seemingly not possible unless the astrophysical parameters can be independently constrained to high precision. On the other hand, keeping the astrophysical parameters open alongside the magnetic and excess radio parameters is expected to lead to mutual degeneracy, a study of which requires scanning of the full parameter space and falls beyond our current scope. As such, we defer these possibilities to future studies.

To wrap things up, this paper is an attempt at a comprehensive analytical study of the bounds on non-helical, nearly scale-invariant PMFs, that are achievable based on pre-reionization 21-cm physics ($z\sim15-25$) via the current EDGES observation and an upcoming 21-cm facility like the SKA-Low. Interesting possibilities in view of PMFs also exist well outside this redshift range towards both higher and lower ends. In particular, the 21-cm signal from the dark ages spanning $z\sim30-100$ is expected to be a pristine source of cosmological information \cite{Mondal:2023xjx}, being free from the astrophysical uncertainties associated with stellar processes at lower redshifts. The prospects of constraining PMFs based on the global signal from the dark ages have recently been explored in Ref. \cite{Mohapatra:2024djd}. On the other hand, EoR and post-EoR physics may provide independent bounds on the PMF parameters, as investigated earlier in Refs. \cite{Schleicher:2011jj,Pandey:2014vga}. As a possible future direction, it is worth re-investigating these scenarios in a joint manner based on our combined analysis, alongside the inclusion of additional current and future datasets, \textit{e.g.} from CMB and baryon acoustic oscillations (BAO), hence forecasting on the synergy of future cosmological missions when it comes to constraining the PMF parameters. We plan to address some of these possibilities in our upcoming works.

\acknowledgments
We thank Antara Dey, Pathikrith Banerjee, Rahul Shah and Sourav Pal for fruitful discussions and computational assistance. We also thank the anonymous referee for their insightful suggestions, which led to substantial improvement of the manuscript. The authors gratefully acknowledge the use of the publicly available codes \texttt{CLASS}~\cite{2011arXiv1104.2932L,2011JCAP...07..034B} and \texttt{CosmicFish}~\cite{Raveri:2016xof,Raveri:2016leq}, alongside the computational facilities of Indian Statistical Institute (ISI), Kolkata. Research work of AB and DP are supported by Senior Research Fellowships respectively from CSIR (File no. 09/0093(13641)/2022-EMR-I) and from ISI kolkata. SP thanks the Department of Science and Technology, Govt. of India for partial support through Grant No. NMICPS/006/MD/2020-21.

\appendix
\section{Discussion on HI and HeI ionization fractions}
\label{app:ionization_fraction}
In this appendix, we discuss the modelling of the ionization fractions of HI and HeI, according to Ref.~\cite{Seager:1999bc}. This leads to the evolution of proton fraction and singly ionized helium fraction which has been shown in Eqs.~\eqref{eq:xp_evol} and \eqref{eq:xHe_evol}, respectively.

In the above set of equations, $\alpha$, Case-B recombination coefficient, has different fitting forms for HI and HeI. For HI, it can be expressed as~\cite{1994MNRAS.268..109H,1991A&A...251..680P}
\begin{equation}
\label{eq:alpha_H}
    \alpha_{\rm H} \equiv F\times 10^{-19}\frac{at^{b}}{1 + ct^{d}} \,\,  \mathrm{m^{3}s^{-1}},
\end{equation}
with $a=4.309$, $b=-0.6166$, $c=0.6703$, $d=0.5300$, $F=1.14$ and
$t\equiv \tk/10^{4}\,$K. On the other hand, for HeI, it is given by~\cite{hummer1998recombination}
\begin{equation}
\label{eq:alpha_He}
    \alpha_{\rm He} \equiv q\left[\sqrt{\tk\over T_2}\left(1+\sqrt{\tk \over T_2}\right)^{1-p}
    \left(1+\sqrt{\tk\over T_1}\right)^{1+p}\right]^{-1}\,
     \mathrm{m^{3}s^{-1}},
\end{equation}
with $q=10^{-16.744}$, $p=0.711$, $T_1=10^{5.114}\,$K, and $T_2=3\,$K. We have denoted the photoionization coefficient by $\beta$ in Eqs.~\eqref{eq:xp_evol} and \eqref{eq:xHe_evol} (suffix carries the usual meaning), which can be expressed in terms of $\alpha$ as~\cite{Seager:1999bc}
\begin{eqnarray}
\label{eq:beta}
    \beta \equiv \alpha\, \left(2\pi m_e k_{\!_{B}} \tk/h_P^2\right)^{3/2}\exp{\left(\frac{-h_P\,\nu_{2s}}{k_{\!_{B}} \tk}\right)},
\end{eqnarray}
with $m_e$ being the electron mass.

The cosmological redshifting of HI Ly-$\alpha$ photons and HeI $2^1 p- 1^1 s$ photons are denoted respectively by $K_{\rm H}\equiv \frac{\lambda^3_{\rm H}}{8\pi\,H(z)}$ and $K_{\rm He}\equiv \frac{\lambda^3_{\rm He}}{8\pi\,H(z)}$~\cite{Seager:1999bc}. As for the specific values of the parameters involved at this point, we have used the atomic data given in Tab.~\ref{tab:atomic_data}~\footnote{The neutral helium fraction can be calculated from the helium abundance $Y_{He} = 0.2449 \pm 0.0040$~\cite{Aver:2015iza} from Big Bang nucleosynthesis (BBN), which results in $f_{He} \approx 0.08$.}.

\begin{table}[h!]
    \centering
    \begin{tabular}{|c|c|c|}
        \hline
        Quantities & Physical meaning & Values\\
        \hline
        \hline
        $\lambda_{H2p}$ & H Ly-$\alpha$ wavelength & $121.5682$ nm\\
        $\lambda_{He}$ & He $2^1p-1^1s$ wavelength & $58.4334$ nm\\
        $\nu_{H2s}$ & H $2s-1p$ frequency & $3.29\times 10^{15}$ Hz\\
        $\nu_{He2^1s}$ & He $2^1s-1^1s$ frequency & $5\times 10^{15}$ Hz\\
        $\Lambda_{\rm H}$ & H $2s-1s$ two photon rate & $8.22458$ s$^{-1}$\\
        $\Lambda_{\rm He}$ & He $2^1s-1^1s$ two photon rate & $51.3$ s$^{-1}$\\
        $f_{\!_{\rm He}}$ & number of helium fraction & $0.08$\\
        \hline
    \end{tabular}
    \caption{List of the atomic data used to model the ionization history, following Ref.~\cite{Seager:1999bc}.}.
    \label{tab:atomic_data}
\end{table}

\section{Interpretation of $B(z)$ in terms of the magnetic power spectrum}
\label{app:bzandpbk}
The physical interpretation of the time-evolving magnetic field $B(z)$ is as follows. It can be described in terms of the magnetic power spectrum $P_B(k)$ in a step-by-step manner, following \emph{e.g.} Sec. 3 of Ref. \cite{Minoda:2018gxj}. First, $P_B(k)$ is given by a power-law in $k$ as
\begin{equation}
    P_B(k)=\dfrac{B_n^2}{k_n^3}\left[\dfrac{(2\pi)^{\nb+5}}{\Gamma\left(\frac{\nb+3}{2}\right)}\right]\left(\dfrac{k}{k_n}\right)^{\nb}\:,
\end{equation}
where $k_n=2\pi$ Mpc\textsuperscript{-1} is conventionally chosen as a reference scale, and $B_n$ is the normalized PMF amplitude on a scale of $1$ Mpc\textsuperscript{-1}. In terms of this power spectrum, the average comoving PMF amplitude at present time smoothed on a length scale $\lambda$ (typically chosen to be $\sim1$ Mpc) is defined to be
\begin{equation}
    B_\lambda^2\equiv\int\dfrac{d^3k}{(2\pi)^3}e^{-k^2\lambda^2}P_B(k)=B_n^2\left(\dfrac{k_\lambda}{k_n}\right)^{\nb+3}\:,
\end{equation}
where $k_\lambda\equiv2\pi/\lambda$ is introduced as the inverse of the smoothing scale. 

Clearly, for a nearly scale-invariant PMF with $\nb\to-3$ as chosen in this work, one roughly ends up with $B_\lambda\approx B_n$. Accordingly, we identify this comoving field strength $B_n$ (which originally appears as the normalization factor in $P_B(k)$) with the present day magnetic field strength denoted by $B_0$ in our work. The redshift-dependence of the smoothed amplitude is then given by $B(z)\equiv B_0(1+z)^2f^{n_B+3/2}(t)$, where the second factor encapsulates the dissipative correction (due to AD and DT) on top of the usual cosmic expansion factor \cite{Minoda:2018gxj}. While the authors of Ref. \cite{Minoda:2018gxj} subsequently work in terms of the function $f(t)$, we switch over to the approach of Ref. \cite{Bera:2020jsg}, whose authors directly solve for the dynamics of $B(z)$, which is more illuminating from a conceptual point of view.

\bibliographystyle{JHEP} 
\bibliography{mybib.bib}

\end{document}